\documentclass[manuscript]{aastex502/aastex}
\input{psfig}
\usepackage{graphicx}
\usepackage{amsmath}
\usepackage{amssymb}

\begin{document}

\title{The dynamics of Jupiter and Saturn in the gaseous proto-planetary disk}

\author{\textbf{\large Alessandro  Morbidelli}}
%\affil{\small\em Observatoire de la C\^ote d'Azur}

\author {\textbf{\large Aur\'elien Crida}}
\affil{\small\em Observatoire de la C\^ote d'Azur}

\author{\hbox{}}

\author{Corresponding author:\\
Alessandro Morbidelli\\
OCA\\
B.P. 4229\\
06304 Nice Cedex 4, France\\
email: morby@obs-nice.fr}

%\newpage

\begin{abstract}

We study the possibility that the mutual interactions between Jupiter
 and Saturn prevented Type II migration from driving these planets
 much closer to the Sun.  Our work extends previous results by Masset
 and Snellgrove (2001), by exploring a wider set of initial conditions
 and disk parameters, and by using a new hydrodynamical code that
 properly describes for the global viscous evolution of the
 disk. Initially both planets migrate towards the Sun, and Saturn's
 migration tends to be faster. As a consequence, they eventually end
 up locked in a mean motion resonance. If this happens in the 2:3
 resonance, the resonant motion is particularly stable, and the gaps
 opened by the planets in the disk may overlap. This causes a drastic
 change in the torque balance for the two planets, which substantially
 slows down the planets' inward migration. If the gap overlap is
 substantial, planet migration may even be stopped or reversed. As the
 widths of the gaps depend on disk viscosity and scale height, this
 mechanism is particularly efficient in low viscosity, cool disks. The
 initial locking of the planets in the 2:3 resonance is a likely
 outcome if Saturn formed at the edge of Jupiter's gap, but also if
 Saturn initially migrated rapidly from further away. We also
 explore the possibility of trapping in other resonances, and the
 subsequent evolutions. We discuss the compatibility of our results
 with the initial conditions adopted in Tsiganis et al. (2005) and
 Gomes et al. (2005) to explain the current orbital architecture of
 the giant planets and the origin of the Late Heavy Bombardment of the
 Moon.

\end{abstract}

\section{Introduction}

The general theory of planet-gas disk interactions (see for instance Lin and
Papaloizou, 1979; Goldreich and Tremaine, 1980; Papaloizou and Lin, 1984; Lin
and Papaloizou, 1986; Ward, 1986, 1997) predicts a systematic migration of the
planets towards the central star.  This prediction received a spectacular
confirmation with the discovery of the first extra-solar planets, on orbits
with semi-major axes comparable to or smaller than those of the terrestrial
planets in our Solar System (the so-called hot and warm Jupiters).

Despite planet migration is undoubtedly a fact, it is not a general rule, or
-at least- not a rule without exceptions. In our Solar System, Jupiter and
Saturn should not have migrated substantially, despite they evidently formed
in a massive gaseous disk in order to accrete their hydrogen rich atmospheres.
In fact, the existence of Uranus and Neptune outside of Saturn's orbit, and of
the Kuiper belt beyond Neptune, constrains the inward migration of Jupiter and
Saturn within a few AUs at most (probably much less; see below). In
extra-solar systems, with the extension of the timescale of observations, new
planets have been found at distances from the parent star comparable to that
of Jupiter. Thus, it is important to study non-generic mechanisms that, in
some cases, might stop planet migration, or slow it down significantly.

A new tight constrain on the orbits of Jupiter and Saturn at the time of
disappearance of the gas disk comes from a recent model developed to explain
the origin of the Late Heavy Bombardment (LHB) of the terrestrial planets
(Gomes et al., 2005). In addition to the LHB, this model explains 
the current orbital architecture of the giant planets (orbital separations, 
eccentricities and mutual inclinations; Tsiganis et al., 2005) the origin
and orbital distribution of the Trojans of Jupiter (Morbidelli et al., 2005)
and the structure of the Kuiper belt (Levison et al., 2007). If true, this
model implies that, at the disappearance of the gas, 
the giant planets and the primordial trans-Neptunian
planetesimal disk were originally in a compact configuration. This limits
severely the radial migration of the planets during the preceeding phase,
dominated by the interactions with the gas. Moreover, the orbits of the
planets were quasi-circular, and the ratio of the orbital periods of Saturn
and Jupiter was smaller than 2. Therefore, it is important to investigate
whether these constraints are consistent with the dynamics of the planets in
the gas disk, as far as we understand it. 

A pioneer work in this direction has been done by Masset and Snellgrove (2001;
MS01 hereafter). In that letter the authors presented the first numerical
simulation of the evolution of Jupiter and Saturn in a gas disk. Saturn
initially started at twice the heliocentric distance of Jupiter. After a phase
of inward runaway migration, Saturn was captured into the 2:3 mean motion
resonance with Jupiter. At that point, the planets reversed their migration,
moving outward in parallel, while preserving their resonant relationship.
This result is interesting for our purposes in two respects. First, it shows
that a two-planet system in some configurations can avoid migration toward the
central star. Second, the stable relative configuration achieved by Saturn and
Jupiter is characterized by a ratio of orbital periods smaller than 2, as
required by the LHB model discussed above.

Therefore, we think that it is timely to investigate more in detail the
mechanism unveiled by MS01, in particular exploring a wider range of
parameters. In fact, there are a few open issues on the validity of the
mechanism and its consistency with the solar system structure that we need to
address:
\begin{itemize}
\item[{\it i})] it is unclear how the
  mechanism depends on the adopted initial conditions. If the initial, inward
  runaway migration of Saturn is a crucial aspect, then the overall result
  might depend on the numerical resolution of the disk and on its initial
  state (namely, whether the planets are dropped into a virgin disk, or they
  are allowed to sculpt the disk for some time before they are let free to
  migrate). In fact, some researchers could not successfully
  reproduce the MS01 simulation (Kley, private communication), possibly
  because of these issues. Moreover, a wide range inward migration of Saturn
  might be inconsistent with the compact orbital architecture of the giant
  planets invoked by the LHB model.
\item[{\it ii})] Jupiter --and to some lesser extent Saturn--, being a
  giant planet, undergoes type II migration. It is well known that
  this kind of migration follows the global evolution of the disk. The
  numerical algorithm used in MS01 could not model the evolution of
  the disk correctly, as it considered only an annulus of it, with
  wise --but nevertheless arbitrarily chosen-- boundary conditions. In
  Crida et al. (2007) we have presented a new hybrid scheme that
  allows the computation of the global evolution of the disk, using a
  system of nested 1D and 2D grids. Consequently, this algorithm
  allows a correct simulation of Type II migration.  It is true that
  MS01 argues that, once the planets are in the 2:3 resonance, they
  are no longer locked in the evolution of the disk (the reason for
  which they can move outward, despite the disk has a global motion
  towards the Sun).  Nevertheless, the torque unbalance that MS01
  measured depends critically on the mass of the disk inside the orbit
  of Jupiter, and the latter depends on the global evolution of the
  disk. In particular, it is known that giant planets can open
  cavities in the inner part of the disk (Rice et al., 2003; Varniere
  et al., 2006; Crida and Morbidelli, 2007). If this happened in this
  case, Jupiter and Saturn could not migrate outward.  Thus, we think
  that it is important to re-simulate the dynamics of Jupiter and
  Saturn using our new, trustable algorithm.
\item[{\it iii})] the evolution of Jupiter presented in MS01 is probably
  inconsistent with the solar system architecture. In fact, after Jupiter and
  Saturn lock in their mutual 2:3 resonance, their outward migration is rather
  fast. Jupiter increases its orbital radius by $\sim 40$\% in 1,000 orbits. 
  If this really occurred in the Solar System, Jupiter would have been at some
  time in the middle of the asteroid belt. The properties of the asteroid
  belt (in particular the quite tight zoning of the taxonomic types) exclude
  this possibility. Thus, we need to find orbital evolutions that are much
  more stationary than the one presented in MS01.
\item[{\it iv})] MS01 claim that the trapping of Jupiter and Saturn into the
  2:3 resonance is the most likely outcome of their evolution in the gas disk.
  This might be problematic in the context of the LHB model (Gomes et al.,
  2005). That model argues that the ratio of Saturn's and Jupiter's orbital
  periods was smaller than 2, but requires that it was not too much smaller
  than this value. Otherwise, the mass of the planetesimal disk remaining
  after the disappearance of the gas would not have been sufficient to drive
  Saturn across the 1:2 resonance with Jupiter, which is required to trigger
  the planetary instability responsible for the origin of the LHB. Thus, it is
  important to investigate if other resonances between Jupiter and Saturn are
  possible in the context of the mechanism of MS01, or if there are
  possibilities to leave the resonance towards the end of the gas disk
  lifetime.  
\end{itemize}

With these goals in mind, this paper is structured as follows. In
Sect.~\ref{MS} we briefly review MS01 result using a very similar simulation
(and the same numerical code), and then we discuss the dependence of the
results on the initial location of Saturn. In Sect.~\ref{dependences} we use
the algorithm of Crida et al. (2007) to investigate how the dynamics of
the Jupiter-Saturn couple depends on the disk's aspect ratio and viscosity,
and also on numerical parameters such as disk's resolution and smoothing
length.  In Sect.~\ref{accrete} we briefly address the effect of the accretion
of mass onto the planets, already partially discussed in MS01. In
Sect.~\ref{masses} we explore the dynamics for different values of the planet
masses, to understand how generic the MS01 mechanism can be for extra-solar
planet cases. In Sect.~\ref{resonance} we discuss possible ways to reconcile
the MS01 mechanism with the LHB model of Gomes et al. Our conclusions are then
recollected in the summary section.

\section{The Masset--Snellgrove mechanism}
\label{MS}

Fig.~\ref{MS-orig} is a quite close reproduction of Fig. 1 in MS01. It has
been obtained using the code presented in Masset (2000a,b) (the same as used
in MS01), with similar parameters. Specifically, Jupiter is initially at
$r=1$, while Saturn is at $r=2$. The disk extends from $r=0.3$ to 5. It is
modeled in two dimensions, using a grid with resolution 282 in radius and 325
in azimuth.  Its initial surface density is uniform in radius and is equal to
$6\times 10^{-4}$ in our units (the mass of the Sun is 1), which corresponds
to the surface density of the minimal mass nebula (Hayashi, 1981) if the unit
of length corresponds to 5 AU. The boundary conditions allow outflow, but not
inflow. We adopt an $\alpha$ prescription for the viscosity (Shakura and
Sunyaev, 1973), with $\alpha=6\times 10^{-3}$, and assume a constant aspect
ratio $H/r=4$\% in the disk's equation of state.

Both Jupiter and Saturn initially migrate inward, as expected. The
migration of Saturn accelerates exponentially, in a runaway --also
called {\it Type III}-- regime (Masset and Papaloizou, 2003) {that
is well fitted by an exponential curve up to $t\sim 100$}.  After a
time of about 100 Jovian initial orbital periods, Saturn crosses the
1:2 mean motion resonance (MMR) with Jupiter, because its migration is
{faster  than the threshold
below which the capture into resonance is certain in the `adiabatic'
approximation (Malhotra, 1993):
$|\dot{a}_S|/(a_S\Omega_S) << 0.5 j (j+1) \mu_J e_S$, for the $j :
j+1$ resonance, where $\mu_J$ is the mass ratio of Jupiter to the
central object, and $a_S$, $e_S$ and $\Omega_S$ are Saturn's
semi-major axis, eccentricity and angular velocity, respectively
(MS01).}  {Once within the 1:2 resonance, Saturn's runaway
migration breaks. This happens because Saturn approaches the outer
edge of Jupiter's gap; thus the disk inside its orbit starts to be
partially depleted and consequently Saturn's coorbital mass deficit 
becomes smaller that the planet's own mass (Masset and Papaloizou, 2003).

After this change in migration regime,} Saturn's inward motion
continues at a slower, approximately constant rate. After 280 initial
orbital period, the migration of Jupiter stops and then it
reverses. As they move in opposite directions, Jupiter and Saturn are
eventually captured in their mutual 2:3 MMR, at $t=350$. After this
event, the two planets migrate outward in parallel. The eccentricity
of Saturn stabilizes around 0.012 and that of Jupiter around 0.003. It
is worth noticing that the ratio of semi major axes of the two planets
does not correspond to a 2:3 ratio of the Keplerian orbital
periods. It rather corresponds to a 5:8 ratio. However, by looking at
the behavior of the resonant angles, we have checked that the planets
are really captured in the 2:3 MMR. The gravity of the disk displaces
the mean motion resonances relative to the unperturbed keplerian
location.

As we said in the introduction, it is an unsolved issue whether the initial
evolution of Saturn, with its runaway migration and fast passage across the
1:2 MMR, plays a role in the subsequent dynamics. For instance, MS01 argue
that the capture in the 2:3 MMR is favored by Saturn's eccentricity being
enhanced during the previous 1:2 MMR crossing. Our understanding of planet
formation is too limited to assess with confidence where Saturn formed.
However, there is an emerging view that Saturn might have accreted its
atmosphere (and therefore acquired the bulk of its mass) when it was already
quite close to Jupiter. In fact, the immediate neighborhood of Jupiter's gap,
being a local maximum of the disk's surface density, acts as an accumulation
point of dust and small planetesimals (Haghighipour and Boss, 2003), and thus
appears as a sweet spot for the growth of Saturn's core. Moreover, Saturn's
core, independently of its formation location, might have suffered Type I
migration until it was halted at the edge of Jupiter's gap, which acts as a
planet trap (Masset et al., 2006), or in a mean motion resonance with Jupiter
(Thommes, 2005). Thus, we believe that it is important to verify whether the
MS01 mechanism can still work if Saturn is released in the proximity of
Jupiter.

Fig.~\ref{MS-proches} shows the result of a simulation that differs from the
previous one only for the initial location of Saturn (now at $r=1.4$). As one
sees, after a short range migration, the planet is trapped into the 2:3 MMR,
and then the evolution is the same as in the previous simulation. Thus, the
migration reversal found in MS01 does not depend on the history of the
previous migration. For this reason, and because of the arguments described
above in favor of a close formation of Saturn, in the following simulations we
will always release Saturn at a distance of 1.4 (which
corresponds to the typical position of the edge of Jupiter's gap). 

\section{Dynamics of Jupiter and Saturn as a function of the disk properties
  and simulation parameters}
\label{dependences}

To explore how the dynamics of Jupiter and Saturn is affected by the main
parameters of the problem, we use, from now on, the numerical scheme described
in Crida et al. (2007). In this scheme, the disk is represented using a system
of 2D and 1D grids. The main portion of the disk, in which the planets evolve,
is represented with a 2D grid in polar coordinates, as usual. The origin of
the coordinates is the barycentre of the system. The inner part of the disk
(ranging from the inner physical radius, e.g. the X-wind truncation radius at
a few tenths of AU, to the inner boundary of the 2D grid) and the outer part
of the disk (ranging from the outer boundary of the 2D grid to the physical
outer edge, e.g. the photo-dissociation radius at hundreds of AU) are
represented with a 1D grid. The 1D grids have open outflow boundaries at the
inner and outer physical edges, and exchange information with the 2D grid for
the definition of realistic, time-dependent boundary conditions of the latter.
The algorithm for the interfacing between the 1D and 2D grids is driven by the
requirement that the angular momentum of the global system (the disk in the 2D
section, plus the disk in the 1D section plus the planet-star system) is
conserved. With this approach, the global viscous evolution of the disk and
the local planet-disk interactions are both well described and the feedback of
one on the other can be properly taken into account. Because the migration of
giant planets depends on the global evolution of the disk, this code provides
more realistic results than the usual algorithms, in which the evolution of the
considered portion of the disk depends crucially on the adopted (arbitrary)
boundary conditions. {For more information and accuracy tests we
  refer the reader to Crida et al. (2007).}

In all simulations presented below, the 2D grid is as before (from 0.3
to 5, with 282x325 resolution). The inner 1D grid starts at $r=0.016$
and the outer 1D grid ends at $r=40$ (which corresponds to about 200
AU in our units). They have the same radial resolution as the 2D
grid. The initial surface density profile of the disk is of type
$\Sigma(r)= 3\times 10^{-4} \exp(-r^2/53)$, as illustrated with a
dash-dotted line in Fig.~\ref{profiles}, and is derived from the
analysis of Guillot and Hueso (2006) for a disk evolving under the
collapse of new matter onto the plane from the proto-stellar cloud,
viscous evolution and photo-evaporation.  The viscosity is assumed
constant, for simplicity (we have verified, as MS01, that an $\alpha$
prescription for the viscosity would not change the results
significantly, as the planets are very close to each other, although
it can affect the global evolution of the disk).

Conversely to what we did in the previous section, we first let the
planets evolve in the disk for 8,000 Jupiter orbits without feeling
the disk's perturbations, assuming an aspect ratio $H/r=3$\% and a
viscosity $\nu=10^{-5.5}$ (in our units, see above). {This
  viscosity at $r=1$ corresponds to $\alpha=3.5\times 10^{-3}$ in a 
Shakura and Sunyaev (1973) prescription.}
This simulation allows the
planets to sculpt the disk, opening gaps around their orbits, and it
sets a new surface density profile of the disk. When we do simulations
with different disk parameters, we start from this profile, and let
the planets evolve for additional 400 Jovian orbits still without
feeling the disk's perturbations, so that the disk profile adapts to
the new situation.  Only at this point we release the planets, letting
them evolve under the effects of the disk and of their mutual
perturbations. This procedure allows us to avoid possible spurious
initial migrations, that might occur if the initial gas distribution
is inconsistent with the presence of the planets.

\subsection{Dependence on the disk aspect ratio}
\label{H/r}

In a first series of runs, we have fixed the value of the viscosity
($\nu=10^{-5.5}$, in our units), and we have studied the evolution of Jupiter
and Saturn as a function of the disk aspect ratio $H/r$.  MS01 found that the
aspect ratio simply changes the outward migration speed, by a quantity
proportional to $(H/r)^{-3}$.  We find that the value of the aspect ratio can
have a much more important impact on the dynamical evolution.

As Fig.~\ref{MS-Hr} shows, if the aspect ratio is small (3 -- 4\%),
the evolution is similar to that previously considered. When released,
Jupiter starts to migrate outward, while Saturn moves inward. After
locking in the mutual 2:3 mean motion resonance, both planets move
outward. In these cases, the migration is indeed faster if the disk is
thinner, as found in MS01.  However, for thicker disks, the evolution
changes qualitatively.  If the aspect ratio is 5\%, we find a
quasi-stationary solution.  After locking in the 2:3 MMR, both Jupiter
and Saturn essentially do not migrate any more. {Actually, Jupiter
moves outward by only 1.5\% in 2000 orbits}.  To our knowledge, this
is the first quasi-stationary solution ever found for a system of giant
planets in a fully evolving disk. If the disk thickness is increased
to 6\%, both planets migrate inward, even after being captured into
the 2:3 MMR. This migration is very slow, compared to that of an
isolated Jupiter in the same disk. This sequence of behaviours
relative to aspect ratio also suggests that, in a flaring disk, the
planets might migrate until they find a position in the disk with the
`good' local aspect ratio that allows them not to migrate any more.

The reason of this parametric dependence of the evolution on $H/r$  
is quite clear if one looks at the gas density
profile at the moment when the planets are released (Fig.~\ref{profiles}). As
explained in Crida et al. (2006), the disk aspect ratio governs the width and
the depth of the gaps opened by the planets.  Therefore, if $H/r$ is
large, there is more gas at the location of Saturn (i.e. just outside
Jupiter's orbit) than in the case where $H/r$ is small.  Consistently, there
is slightly less gas inside of Jupiter's orbit (for $r<0.7$), because less
material has been removed from the common gap formed by the two planets. As
MS01 correctly pointed out, the direction of migration of Jupiter depends on
the balance of the torques that the planet receives respectively from the disk
inside its orbit (which pushes the planet outward) and and from the disk
outside its orbit (which pushes the planet inward).  In the case of an
isolated planet, the torque from the outer disk is typically stronger, so that
the planet migrates toward the Sun. But in this case, because the presence of
Saturn depletes partially the outer disk, this torque is weakened. Obviously,
it is weakened more if the gap at Saturn's position is deeper, namely if the
disk aspect ratio is smaller, as visible in Fig.~\ref{profiles}. Thus, if the
aspect ratio is small enough, the torque received by Jupiter from the inner
disk dominates that from the outer disk, and the planet migrates outward,
feeling a net positive torque. Indeed, this is what we see happening in
Fig.~\ref{MS-Hr}. 

The direction of migration of Jupiter determines the subsequent evolution of
both planets, once they are locked in resonance. The planets have to move in
parallel to preserve the resonant configuration. Therefore there is a
competition between the net positive torque received by Jupiter and the net
negative torque received by Saturn from the disk. Because these torques are
monotonic functions of the planets' masses, and Jupiter is 3 times heavier than
Saturn, in general the positive torque received by Jupiter dominates and the
two planets move outward. In the case with $H/r=5$\%, however, the net torque
felt by Jupiter is close to zero, due to the specific density profile of the
disk, and can be effectively canceled out by Saturn's torque. Thus, a
non-migrating evolution is achieved after the planets lock in resonance.

For completeness and sake of clarity, in the remaining part of this
sub-section we elaborate on some considerations already reported in
sect. 2.4 of MS01.  The principle of Type II migration is that, once a
planet opens a gap, it positions itself inside the gap in order to
balance the torques received from the inner and the outer parts of the
disk. Then, locked into this equilibrium position, it is forced to
follow the slow, global viscous evolution of the disk (Lin and
Papaloizou, 1986), the latter described by the equations in
Lynden-Bell and Pringle (1974).  One could expect that the
Jupiter-Saturn system should evolve in the same way. The outer
migration of the pair of planets should approach Saturn to the outer
edge of its gap, until Saturn feels a stronger torque that
counterbalances the one received by Jupiter. In this situation the
outward migration should stop, and the two planets should start to
evolve towards the Sun, together with the disk.  This, apparently,
does not happen. {For the disk parameters that we explore in this
work,} Saturn is not massive enough to open a clean gap (see
Fig.~\ref{profiles}). Thus, {the conditions for a proper Type~II
migration are never fulfilled (see Crida and Morbidelli, 2007, for a
discussion on Type~II migration)}. If Saturn's radial migration is not
the same as the natural radial motion of the gas, new material flows
into its gap.  However, the gaps of Jupiter and Saturn overlap, so
that material flowing from the outer disk into the coorbital region of
Saturn, after experiencing half of a horse-shoe trajectory relative to
Saturn, can also perform half of a horse-shoe trajectory relative to
Jupiter. The net result is a flow of matter from the outer part of the
disk, through the Jupiter-Saturn common gap, into the inner part of
the disk.  To illustrate this process, Fig.~\ref{profiles_t} shows the
surface density profile of the disk in the simulation with $H/r=3$\%,
at various times. Notice how, in first approximation, the
Jupiter-Saturn gap simply `shifts' through the disk. As the planets
move outward, the disk is rebuilt inside the orbit of Jupiter and the
surface density at the bottom of the gap increases as well. Both
features are diagnostic of a mass flow through the planets system. In
fact, in the code of Crida et al. (2007) that we use, the boundary
conditions cannot act as a source of mass. Thus, an increase of the
surface density in the inner part of the disk is possible only if
there is an influx of mass from the outer disk.

The flow of gas has several effects. First, it unlocks the planets from the
disk, allowing them to move against the gas stream. Second, it has positive
feedbacks on the outer migration of the planets by (i) exerting a corotation
torque on them, as it passes through their horseshoe regions and (ii)
refurbishing the inner part of the disk, which exerts the positive torque on
Jupiter discussed above. The signature of this
feedback is well visible in the simulation with $H/r=3$\% in Fig.~\ref{MS-Hr}: 
the outward migration rate accelerates exponentially, which implies that there
is a positive net torque that increases with the migration speed. 

At this point, one might wonder whether the motion of the planets is dominated
by the torque felt by Jupiter from the inner disk, or by the corotation torque
exerted by the gas flowing through the planets' orbits. The flow of gas
through the orbits of Saturn and Jupiter is the same; the size of the
horseshoe regions of the two planets (and hence the magnitude of the
corotation torque felt by each planet) is proportional to $M_p^\gamma$ for
some $\gamma < 1$; thus, the effect on the migration rate $\dot a_p$ of the
planet is proportional to $M_p^{(\gamma-1)}$, namely it is larger for a
lighter planet. So, if the corotation torque dominated the evolution of the
planets, Saturn would be extracted from the resonance and would migrate away
from Jupiter. As long as this does not happen (as in Fig.~\ref{MS-Hr}), the
corotation torque cannot be the dominant force driving the planets' migration.

  For a further {\bf indication} that the corotation torque {\bf is 
  weaker} than the torque felt by Jupiter from the inner disk
  (Lindblad torque), we have done another simulation, still with
  $H/r=3$\% and $\nu=10^{-5.5}$, but with an initial gas density reduced
  by a factor of 2. The Lindblad torque scales with the gas
  density. Conversely, the corotation torque {\bf does not scale
  simply with the gas density because a component of it depends on the
  the radial speed of the planet relative to the gas, which in turn
  also depends on the gas density} (Masset and Papaloizou, 2003).
  Thus, if the Lindblad torque dominates, we expect the planets to
  have the same evolution, just a factor of 2 slower. If the
  corotation torque dominates, the change in the dynamics can not be
  {\bf trivially} reduced to a simple scaling on time.
  Fig.~\ref{smalldisk} shows the result. The black curves show the
  evolution of Jupiter and Saturn in the nominal gas disk. The grey
  curves show the evolution of the planets in a disk with half the
  initial density. In plotting this second pair of curves, the
  time-span measured relative to the release time (400 orbits) has
  been divided by two.  The black and grey curves superpose almost
  perfectly. This {\bf suggests that the
  Lindblad torque is stronger than the corotation torque}.

\subsection{Dependence on the simulation's technical
  parameters}
\label{tech_params}

Before proceeding further with our exploration of the dynamical evolution of
Jupiter and Saturn, we check the impact of some technical
parameters used in the simulation: specifically the smoothing length for the
gravitational potential and the grid resolution for the disk. 

As usual in 2 dimensional hydro-dynamical simulations, the equations
of motions are regularized in the vicinity of the planet by modifying
the gravitational potential energy $U(\Delta)=-(M_p m)/\Delta$ into
$U_\rho(\Delta)=-(M_p m)/\sqrt{\Delta^2+\rho^2}$, where $\Delta$ is
the distance between the planet of mass $M_p$ and a fluid element of
mass $m$, and $\rho$ is called the {\it smoothing length}. The choice
of an appropriate value for the smoothing length is the subject of a
vast debate. Essentially, two recipes are used: either $\rho$ is
chosen proportionally to the Hill radius of the planet, or
proportionally to the local thickness of the disk.  

In the simulations presented above, our choice of $\rho$ was equal to
60\% of the planet's Hill radius $R_H=a_p(M_p/3)^{1/3}$, where $a_p$
is the semi major axis of the planet and $M_p$ is normalized relative
to the mass of the star.  We have decided to redo the simulation with
$H/r=3$\%, adopting $\rho=0.7 H$, where $H$ is the thickness of the
disk at the distance of the planet (namely 0.03~$a_p$).  In this case,
the new value of $\rho$ for Saturn and Jupiter is, respectively, 73\%
and 50\% of those previously adopted. The new simulation is compared
with the previous one in Fig.~\ref{comp-tch}. Because the simulations
do not start exactly in the same way, for a more meaningful comparison
we have plotted the evolution of the planets only from the time $T_0$
at which Saturn starts its outward migration (this time is slightly
different in the two simulations) and we have renormalized the semi
major axes of the planets by the semi major axis of Jupiter at
$T_0$. As one sees, the difference is not very big. Using the new
value of $\rho$ leads to a slightly faster migration. The reason is
that Saturn opens a wider and deeper gap in the new simulation,
because the smaller value of the smoothing length is equivalent to an
enhancement of its gravitational potential. As we have seen before, a
deeper gap at Saturn's location increases the unbalance of the torques
exerted on Jupiter from the inner and the outer parts of the disk, and
hence leads to a faster outward migration speed. We have performed all
the simulations of Fig.~\ref{MS-Hr} with the new prescription of the
smoothing length.  None of the simulations changes significantly. In
particular we still find a quasi-stationary, non-migrating evolution
in the case with $H/r=5$\% {(Jupiter now migrates outward only
by 0.5\% in 2,000 orbits)}, and an inward migration in the case of
$H/r=6$\%. Because the choice of $\rho$ based from the local thickness
of the disk seems to us somewhat more physically motivated, we will adopt this
prescription in all the simulations presented further in this paper.

{Whatever the choice of $\rho$ above ($0.6 R_H$ or $0.7 H$), the
  smoothing length is always a big fraction of the planet's Hill
  radius. Thus, it is interesting to explore what would happen if we chose a
  much shorter smoothing length. In Fig.~\ref{short-smooth} we compare
  the simulation with $H/r=$4\% and $\rho=0.6 R_H$ (black curves, already
  illustrated in Fig.~\ref{MS-Hr}) with one with the same disk
  parameters, but $\rho=0.25 R_H$ (grey curves) and one with the same
  prescription of $\rho$ but where we have nullified the torques
  exerted on the planet(s) by the gas inside their respective Hill
  spheres (light grey curves).  The exclusion of the torques from the
  regions neighboring the planets is never implemented in all other
  simulations presented in this paper. As one sees, the planets'
  migration rates depend quite strongly on the adopted prescription
  for smoothing and torque calculation. This is because, if $\rho$ is
  small, Saturn opens a deeper gap at its location, which enhances the
  unbalance of the torques felt by Jupiter. What is important,
  however, is that in all cases the migration
  is {\it outward}. This, once again, shows the robustness of the MS01
  mechanism.}

The resolution of the grid used to represent the disk can also have, in
principle, an important impact on the evolution of the system. In particular
it can affect the corotation torque that, as we have seen, plays a role in the
outward migration of the planets. To test the effects of the grid resolution,
we have repeated the simulation with $H/r=3$\% and $\rho= 0.7 H$, increasing
by a factor of two both the radial and azimuthal resolutions of the 2D grid,
and the radial resolution of the 1D grids. The new simulation is also plotted
in Fig.~\ref{comp-tch}. As one sees, the difference with respect to the
simulation with our nominal resolution is negligible. Given the computational
cost of the high resolution simulation, we will continue to use 282$\times$325
cells in the 2D grid in the subsequent experiments.

{Another technical issue concerns the initial condition for the
  gas distribution. As we said above, we start from a gas profile
  carved by the planets on fixed orbits in a simulation spanning 8,000
  periods at $r=1$, with $H/r$=3\%, $\nu=10^{-5.5}$ and $\rho=0.6
  R_H$. However, when we change the parameters of the simulation, we
  only wait for additional 400 orbital periods before letting the
  planets free to evolve. This second delay might not be long enough
  for the gas to respond to the new conditions, possibly introducing
  artefacts in the subsequent planet dynamics. To check if this is
  indeed the case, we plot in Fig.~\ref{release} two simulations, for
  $H/r=$4\%,$\nu=10^{-5.5}$ and $\rho=0.7 H$. In one simulation (black
  curves) the planets have been released after 400 orbits, as usual;
  in the second simulation, the planets have been released after 5,000
  orbits. The evolutions after the release time match so perfectly
  that, in order to see the two sets of curves we had to downshift the
  gray ones by 1\%! Thus, we conclude that our relatively short
  rlelease time of 400 orbits does not introduce significant
  artefacts. }

\subsection{Dependence on the disk viscosity}
\label{nu}
 
We have done a series of simulations, changing the value of the viscosity,
from $\nu=10^{-6}$ to $2\times 10^{-5}$ in our units. The disk aspect ratio is
4\% in all simulations. {Thus, at $r=1$, these viscosities
  correspond to $\alpha$ ranging from $6.25\times10^{-4}$ to
  $1.25\times 10^{-2}$, in a Shakura and Sunyaev (1973) prescription.} 
As usual, Saturn starts at $r=1.4$ and Jupiter at
$r=1$. The results are illustrated in Fig.~\ref{MS-nu}.

%Given that the viscosity, like the aspect ratio, governs the width and the
%depth of the gaps (Crida et al., 2006), we could expect that the dependence
%of the dynamics on the considered parameter is analog to the one that we have
%illustrated in sect.~\ref{H/r}: a fast outward migration for low viscosity
%(i.e. deep and large overlapping gaps), a quasi-stationary solution for some intermediate
%value, and an inward migration for large viscosity (i.e. shallow and narrow,
%non-overlapping gaps). The reality, however, is not that simple.

For a viscosity $\nu=10^{-6}$, Jupiter migrates outward after it has been
released. Saturn initially migrates inward and, after being locked in the 2:3
MMR with Jupiter, the two planets migrate outward in parallel. 
For a viscosity $\nu=5\times 10^{-6}$, the evolution is qualitatively
similar. The outward migration speed, however, is faster than in the previous
case.
%, in contrast with what we would expect from the analogy with
%sect.~\ref{H/r}. 
The reason is that the inner edge of Jupiter's gap is further away
from the planet in the $\nu=10^{-6}$ case than in the $\nu=5\times 10^{-6}$
case (see Fig.~\ref{profiles2}), so that the positive torque felt from
the inner disk is weaker in the first case.  
 
For a viscosity $\nu=10^{-5}$, Saturn --when released-- has some
erratic motion, which is slightly outward, on average, until
$T=1200$. During this time-span, Jupiter, which is also migrating
outward, approaches Saturn.  Eventually Saturn has a short inward
migration and is captured in the 2:3 MMR with Jupiter, and the two
planets migrate outward together. Their common outward migration is
slower than in the previous cases. If the viscosity is increased to
$2\times 10^{-5}$, as soon as released Saturn migrates outward.
Jupiter in the meantime migrates inward. The mutual 1:2 MMR is crossed
at $T=880$. The eccentricity enhancement that results from this
resonance crossing breaks Saturn's outward migration. The planet
starts a `normal' inward migration, at a rate comparable to that of
Jupiter. The two planets are close to the 1:2 MMR, but not locked in
it. The resonant angles are in fact in circulation. The reason for the
initial behavior of Saturn in these two simulations is most likely due
to the corotation torque. As Fig.~\ref{profiles2} illustrates, for
these values of the viscosity there is quite a large amount of gas at
Saturn's location, and the outer edge of Saturn's gap is very close to
the planet. More importantly, Jupiter's gap becomes shallower. This
reveals that, even before that the planets are released, there is {an important} flow of gas from the outer part of the disk, through the
planetary orbits, towards the inner part of the disk. This flow exerts
a corotation torque on each planet, which, as we discussed above, has
stronger effects on Saturn.

Putting together these results with those of sect.~\ref{H/r}, we
conclude that the mechanism of MS01 works for a large range of values
of aspect ratio and viscosity of the disk. Whenever the disk is enough
thin and of low viscosity, Jupiter and Saturn can have a common
outward migration, once locked in the 2:3 MMR.  Finding a
quasi-stationary solution, however, is more delicate. If the disk is
relatively thick (5\% and, presumably, more), a quasi-stationary
solution can be found for some value of the viscosity. Conversely, if
the disk is thin (4\% or, presumably, less), a quasi-stationary
solution may not be found.  The reason is that, if the disk's aspect
ratio is decreased, in principle the viscosity needs to be increased
in order to maintain a density at Saturn's location that is
sufficiently large to exert on Jupiter a torque that counter-balances
the one received by the planet from the inner disk. This larger
viscosity, however, tends to destabilize Saturn, because it generates
a stronger flow that exerts an more important corotation torque on the
planet.

\section{The effect of mass accretion onto the planets}
\label{accrete}

In all previous simulations, the mass of the planets was kept constant
with time. {Lubow et al. (1999) and Kley (1999) 
showed that the accretion of mass
by Jovian or sub-Jovian planets is non negligible for most values of
the disk's parameters.}  The investigation of the effects of mass
accretion onto the planets is {therefore} interesting. Mass
accretion exerts additional torques onto the planets and breaks the
flow of the gas across the planetary orbits. So, in principle it could
modify the dynamics significantly.

MS01 already explored the effect of mass accretion onto Jupiter, and found
that it is negligible even from the quantitative point of view. Here we
consider also the effect of mass accretion onto Saturn, which might have a
larger impact on the dynamics. Our understanding on how planet accretion
proceeds, and how it stops, is still too vague to be able to assert a priori
which planet should have had a more important mass growth rate. 

As MS01, we have implemented mass accretion onto the planets following
the recipe of Kley (1999). It consists in removing a fraction of the
material in the Hill sphere of the planet and adding it to the mass of
the planet.  The amount which is removed in one {\bf time-unit} is
imposed as an input parameter.  More specifically, we apply the input
removal rate in the inner Hill sphere (extended up to 45\% of the Hill
radius $R_H$); we apply 2/3 of the removal rate in the region from
0.45 to 0.75~$R_H$ and no removal rate in the region beyond
0.75~$R_H$. We have done 6 simulations, with three removal rates
applied to Saturn only or both Jupiter and Saturn.  The removal rates,
(expressed as fraction of mass removed in the unit of time, which is
$1/2\pi$ of the initial Jupiter's orbital period) are 0.1, 1 and 5, as
in Kley (1999).  All the simulations started from an intermediate
state achieved in the simulation with $H/r=3$\%, $\nu=10^{-5.5}$
{(equivalent to $\alpha=3.5\times 10^{-3}$)} and $\rho=0.7 H$ (already
presented in Fig.~\ref{comp-tch}), precisely after a time
corresponding to 900 initial Jovian orbital periods after the release
of the planets.
 
Figure~\ref{MS-accr} shows the result in the case of an accretion rate of 1
applied on Saturn only and compares it with the nominal simulation that we
started from, where no accretion was allowed. We notice that the outward
migration rate of Saturn and Jupiter (up to 1700 orbital periods) is
significantly smaller. During this time, the eccentricity of Saturn is larger
than in the case without accretion ($\sim 0.05$ instead of $\sim 0.02$), which
means that Saturn is offering a stronger resistance to the outward push
exerted by Jupiter through the 2:3 MMR. This is most likely due to the fact
that Saturn is growing in mass, so that the negative torque that it receives
from the outer part of the disk increases. In fact, from $t=900$ to $t=1700$,
Saturn doubles it mass, in an essentially linear mode.

At $t=1700$ the dynamical evolution changes abruptly. The mass {  growth of Saturn is accelerated, so that the planet reaches} one
  Jupiter mass at $t=1800$.  This abrupt flow of mass onto the planet,
  essentially from the outer disk, exerts a strong positive
  torque. Therefore Saturn is extracted from the 2:3 MMR with Jupiter
  and runs away from it. {Once separated from Jupiter, `Saturn'
  starts an inward, Type~II-like migration, despite of the positive
  torque due to the accretion of gas, which is still ongoing.}

The simulation where the mass of Jupiter is also allowed to grow, is
essentially identical to the one presented in Fig.~\ref{MS-accr}. We note in
passing that during the linear growth regime, while the mass of Saturn
doubles, the mass of Jupiter increases by only 15\%. This shows that
neglecting the growth of Saturn while allowing the growth of Jupiter is not
justified. We also remark that the growth of the planets does not stall until
they reach a mass of several Jupiter masses. This stresses the unsolved
problem of explaining the final masses of the giant planets of the 
solar system (and of extra-solar systems in general).

The simulations with a smaller (0.1) or larger (5) removal rate parameter
behave essentially like that presented in Fig.~\ref{MS-accr}. Obviously,
during the linear mass growth regime, the deviation with respect to the
nominal simulation without mass accretion is smaller in the first case 
and larger in the second case. Even in the case with a removal rate of 5,
though, we observe an outer migration of Jupiter of Saturn. This implies
that this kind of dynamical evolution is robust with respect to the accretion
rate, {unless the latter is very high.} 

\section{Generic two-planet dynamics: dependence on the individual masses and
  mass ratio} 
\label{masses} 

Although this paper is devoted to the evolution of Jupiter and Saturn, it is
interesting to do a quick exploration of how the dynamics changes with the
masses of the planets. We have done three simulations, all with $H/r=5$\%,
$\nu=10^{-5.5}$ {(corresponding to $\alpha=1.25\times 10^{-3}$)}
and $\rho=0.7 H$ (these parameters corresponds to the
quasi-stationary solution for the Jupiter-Saturn system): the first one
assumes that the inner planet has the mass of Saturn and the outer one has the
mass of Jupiter; the second one assumes both masses are equal to one Jupiter
mass; the third simulation multiplies the masses of the real planets by a
factor of three.

The first two simulations give no surprises. As we explained in
sect.~\ref{H/r}, the outward migration is possible only if the inner planet is
more massive than the outer one. Otherwise the balance between the positive
torque felt by the inner planet and the negative torque felt by the outer
planet is in favor of the latter one. In fact, in both the first and the
second simulation the planets migrate inward. Initially, the outer planet
migrates faster than the inner one, so that the two planets get captured in
the 2:3 MMR after some time.

The third simulation is the most interesting. In this case the mass ratio is
the same as in the Jupiter-Saturn case, favoring an outward migration.
However, because the outer planet is more massive than Saturn, it may be more
difficult to unlock the evolution of the planetary system from the evolution
of the gas, which favors an inward migration.  So, the result of this
experiment is not evident a priori. Fig.~\ref{3J-J} shows the outcome. When
the planets are released (as usual after 400 orbital periods of the inner
planet), the inner planet starts to move outward as expected. The outer planet
remains essentially on the spot. Before that a resonant configuration is
achieved, the planets destabilize each other, because their separation
corresponds to less than 3 mutual Hill radii (a mutual Hill radius is defined
as $[(a_1+a_2)/2] [(M_1+M_2)/3]^{1/3}$, where $a_1, a_2$ are the semi major
axes and $M_1, M_2$ the masses). As a result of this instability, at $t=600$
the outer planet is propelled outward on an orbit with eccentricity equal to
0.25, and the inner planet is kicked inward, onto an orbit with eccentricity
equal to 0.1. Because of the large masses of the planets, the two gaps still
partially overlap.  Therefore, the inner planet feels a net positive torque,
and the outer planet a net inward torque and, at $t=700$--800, they start to
migrate in converging directions. During this time, their orbital
eccentricities are damped down to less than 0.05. At $t=1000$ the planets are
captured in their mutual 1:2 MMR.  For a while after the resonant capture, the
two planets move outward, but then eventually they stop, in a sort of
quasi-stationary configuration. Our interpretation is that the gap of the
outer planet is much more impermeable to the gas flow than in Saturn's case.
Consequently, under the push felt from the inner planet, the outer planet
simply approaches the edge of its gap {and modifies its profile} until the torque that it receives from
the outer disk can counterbalance the torque from the inner planet. This stops
the migration.

In conclusion, the mechanism proposed in MS01 is not necessarily specific to
our solar system. It can apply to extra-solar planetary systems but only at
given, stringent conditions: (i) the outer planet has to be significantly less
massive than the inner one and (ii) the planets have to be locked in a
resonance characterized by an orbital separation that is sufficiently small to
allow the overlapping of the respective gaps. All of the 20 multi-planets
extra-solar systems discovered so far should have suffered a wide range
migration, as suggested by the close proximity of the planets to the central
star (typically, the inner planet is within 1.5--2 AU and the outer planet
within 4 AU). So, we should expect that the mechanism of MS01 did not work in
these systems. In fact, in 13 cases criterion (i) is not fulfilled. In the
remaining cases the planets are too separated, with ratios of orbital periods
larger than 3, so that it is unlikely that they have ever been locked in
resonances with small orbital separation in the past. We predict that
extra-solar systems satisfying both conditions (i) and (ii) will be discovered
in the future, when the observation time-span will become long enough to allow
the detection of distant planets that did not migrate significantly.

\section{Possible ratios of orbital periods of Jupiter and Saturn}
\label{resonance}

In all the simulations reported above, as well as in those of MS01,
whenever Jupiter and Saturn are in a configuration that prevents their
migration towards the Sun, they are locked in the 2:3 MMR. This
supports the idea, proposed in Tsiganis et al. (2005) and Gomes et
al. (2005), that - at the end of the gas disk phase- the system of the
giant planets in our solar system was very compact (i.e. characterized
by small separations between the planets' orbits).  However, from the
quantitative point of view, our results do not support directly the
initial conditions adopted in Tsiganis et al. and Gomes et al. The
initial ratio between the orbital periods of Saturn and Jupiter in
that model was 1.8--1.9. The exact value is not important, but it is
required that it is close to 2, so that Saturn can cross the 1:2 MMR
with Jupiter in $\sim 650$~My (the time of the Late Heavy Bombardment)
due to its interaction with the remaining planetesimal disk. If, at
the end of the gas disk phase, the ratio of orbital periods had been
close to 1.5, it is unlikely that this would have happened (unless a
very massive planetesimal disk is assumed, but this would lead to
other problems concerning the evolution of Uranus and
Neptune). Therefore, in this section we explore different ways to
reconcile the MS01 mechanism with the initial conditions of the LHB
model.

In principle, it is not necessary that Saturn and Jupiter are locked in the
2:3 MMR in order to prevent their inward migration. Other resonances,
characterized by a larger ratio of orbital periods, may work, provided that
the gaps formed in the disk by the two planets are wide enough to overlap.
This would give a constraint on the maximal viscosity and scale height of the
disk for each chosen resonant configuration. Reality, however, is not that
simple, because the resonances located between the 2:3 and 1:2 MMR are much
thinner than first order resonances and they may be characterized by unstable
motion. So, the possibility of capture and permanence of the planets in these
resonances is not guaranteed, a priori.

We have done a series of 5 simulations, starting Saturn at a distance of 1.5
(Jupiter being initially at 1, as usual, so that the initial ratio of the
orbital periods is 1.84), in disks with aspect ratio of 3.5\% and viscosities
in the range 1--3$\times 10^{-6}$. In all simulations we have observed only
captures in the 3:5 MMR, which led to a quasi-stationary evolution or a slow
outward migration of the giant planets. However, in all cases, once captured
in the resonance, the eccentricity of Saturn grew above 0.1 in about 150
initial Jovian orbital periods. This led to an instability of the planetary
motion, which eventually led to a phase of violent scattering among the
planets.  Thus, we conclude that resonances of order larger than 1, located in
between the 2:3 and 1:2 MMR are not viable for a long phase of quiescent,
non-migrating evolution. They either don't capture the planets, or lead to an
unstable motion after a short timescale.

We have also done two simulations with Saturn initially at a distance of $\sim
1.65$ (initial ratio with Jupiter's orbital period of $\sim 2.1$), in a disk
with $H/r=3$\% and viscosity of $5\times 10^{-6}$. In both cases we have
obtained capture in the 1:2 MMR, and a subsequent quasi-stationary evolution
of the semi major axes of the two planets. This shows that the passage across
the resonance without capture observed in MS01 (and in Fig.~\ref{MS-orig}
above) was due to the fact that Saturn was migrating very fast. Our initial
conditions and the low viscosity of the disk allow a slower migration and a
more gradual growth of the eccentricity, which favor capture.  Once captured
in the resonance, despite the eccentricities of the planets are not
negligible, the orbital evolution of the planets looks stable.  We have not
found any obvious way of extracting the planets from the resonance after some
time, and delivering them on orbits with orbital period ratio smaller than 2.
So, we doubt that a capture in the 1:2 MMR during the gas disk phase may be
compatible with the initial conditions of the LHB model.

Finally, we have studied the possibility that Saturn is extracted from the 2:3
MMR with Jupiter, after a long phase of quasi-stationary evolution, and is
transported to larger semi major axis, approaching the 1:2 MMR. 

A first idea is that, as the surface density of the disk decreases during the
disk dissipation phase, the planetary motion might become unstable so that the
planets push each other onto more widely separated orbits. We have rapidly
discarded this possibility, because a `stability
map' shows that the Jupiter-Saturn system at low eccentricity is stable if the
ratio of the orbital periods is larger than 1.45 (Gayon, private
communication). 
 
A second idea is suggested by the simulation presented in
Fig.~\ref{MS-accr}.  The simulation should be considered only at a
qualitative level for several reasons: accretion was started when
Saturn already had one Saturn's mass, so that the final mass of the
planet is larger than the real one; the prescription used for mass
accretion was ad-hoc and idealized. Nevertheless, the simulation shows
that {rapid} accretion of mass onto a planet exerts a {positive}
torque that can extract the planet from the resonance. For instance,
in Fig.~\ref{MS-accr} the ratio of orbital periods of Saturn and
Jupiter at $t-T_0 = 1800$ is 1.85, consistent with the initial
conditions of the LHB model.  Of course, once the planets are
extracted from the resonance, their orbital evolution is no longer at
equilibrium, and migration is resumed. Thus, to advocate a final
position of the planets close to the 1:2 MMR, one has to assume that
the disk disappeared `at the right time'.  Our understanding of
planetary growth is still too poor to draw definite
conclusions. However, the moderate mass of Saturn may be an indication
that its {rapid} growth was indeed aborted by the disappearance of the
disk (Pollack et al., 1996). {Notice  however that, if the growth
  of Saturn is really as rapid as the simulation shows (0.4 Jupiter
  masses in 700 orbits), the nebula has
  to dissipate on a timescale of $\sim 10,000$~y,
  otherwise Saturn would have become too massive.} 

A further, possibly more promising idea, concerns the evolution of the
viscosity of the disk. As we have seen in sect.~\ref{nu}, the MS01 mechanism
works only if the viscosity is sufficiently small. If the viscosity exceeds
some value, Saturn can be extracted from the resonance in a runaway migration
mode (see Fig.~\ref{MS-nu}). Thus, it is interesting to explore the dynamics
of Jupiter and Saturn in the case of a disk whose viscosity increases with
time. In principle, there are a few reasons to believe that the disk's
viscosity might grow towards the end of the disk's lifetime. If the origin of
viscosity is MHD turbulence (Balbus and Hawley, 1991), the viscosity depends
on the ionization of the disk. A sufficiently massive disk is optically thick,
so that the radiation from the star cannot penetrate in the disk and the gas
is not ionized. Thus, a {\it dead zone} can exist inside the disk, at a
typical distance from a few to a few tens of AU, where MRI turbulence is not
sustained and therefore the viscosity is very small (Gammie, 1996). The giant
planets might very well have formed in such a dead zone.  When the disk starts
to disappear, the radiation of the star can penetrate deeper into the disk,
ionizing the disk on the mid-plane at larger heliocentric distance. The dead
zone is re-activated, which causes an important enhancement of the local
viscosity. Moreover, dust tends to have chemical bonds with the ions,
subtracting them from the gas.  Thus, even if undergoing the ionizing effect
of the stellar radiation, a disk might not exhibit MHD turbulence if a
sufficient amount of dust is present (Ilgner and Nelson, 2006a,b). As time
passes, most of the dust is accreted in planetesimals, and therefore cannot
subtract ions as efficiently as before. Therefore, a late disk should be
increasingly coupled to the magnetic field and be characterized by a more
violent turbulence and stronger viscosity.

Motivated by these considerations, we have designed the following experiment.
We considered the simulation with Jupiter initially at $r=1$, Saturn at
$r=1.4$ and a disk with $H/r=5$\% and $\nu=10^{-5.5}$ {(corresponding to $\alpha=1.25\times 10^{-3}$)}, performed assuming a
smoothing length $\rho=0.7 H$. In this simulation, after capture in the mutual
2:3 MMR, Jupiter and Saturn exhibit a remarkable quasi-stationary solution (see
Fig.~\ref{varnu} up to $t=3200$). At $t=3200$ we started to increase the
viscosity of the disk, at the rate of $10^{-9}$ per unit of time (we remind
that in our units the orbital period at $r=1$ is $2\pi$). This rate is totally
arbitrary, and not justified by any astrophysical considerations. As a
consequence of the increase in viscosity, the gaps formed by the planets
become narrower and overlap more marginally. Thus, the shape of the gap formed
by Jupiter becomes more symmetric with respect to the position of the planet,
so that the torque received by Jupiter from the outer part of the disk starts
to dominate over that from the inner part of the disk. Consequently Jupiter
starts to migrate towards the Sun. The migration rate increases with
increasing viscosity. As the resonance with Jupiter moves inward, Saturn also
migrates towards the Sun, but at a smaller rate. In fact, the flow of gas from
the outer disk towards Jupiter exerts a corotation torque on Saturn, slowing
down its inward migration. This extracts Saturn from the 2:3 MMR. As the
viscosity increases, the corotation torque becomes stronger, and eventually
Saturn starts an outward runaway migration.  At $t=7200$ Saturn is very close
to the 1:2 MMR with Jupiter, as required in the initial conditions of the LHB
model. The eccentricities of Jupiter and Saturn are very low, less than 0.005
and 0.01 respectively, which is also consistent with the LHB model. At that
time, the viscosity of the disk is $\nu=3.1\times 10^{-5}$. Given that the
aspect ratio is 5\%, this viscosity would correspond to a value of $\alpha\sim
10^{-2}$, which is still reasonable. Obviously, to support the initial
conditions of the LHB model, one has to assume that the disk disappears at
that time. If this were not the case, and the viscosity kept growing, Saturn
would cross the 1:2 MMR with Jupiter.  Notice that, overall, Jupiter has an
inward migration that covers only 20\% of its initial heliocentric distance.
Thus, in this scenario, to justify its current position, Jupiter should have
formed at about 6.5 AU (and Saturn at about 8.5 AU, to end up, more or less,
at the same position). These ranges of migration are moderate, and do not
violate, a priori, any of the constraints imposed by the current solar system
architecture.

Again, we think that this simulation should be considered only at a
qualitative level. Our knowledge of the evolution of the disk close to its
disappearance is too approximated to be able to build a realistic simulation.
Fig.~\ref{varnu} is presented simply to show that it is possible, in
principle, to release the planets on non-resonant orbits after that they have
spent most of the disk lifetime on resonant, non-migrating ones. Obviously,
making the bridge between the formation of the planets, their dynamics in the
gas disk, and their subsequent evolution in the planetesimal disk remains an
open, crucial problem that goes beyond the scopes of this work.

\section{Summary}
\label{end}

In this paper we have analyzed in detail, by performing many numerical
simulations, the mechanism proposed by Masset and Snellgrove (2001) to
explain why Jupiter and Saturn did not migrate towards the Sun. The
simulations have been done with a new simulation scheme (Crida et al.,
2007), that is particularly suitable to study the migration of the
giant planets. We confirmed that, if Jupiter and Saturn are locked
into their mutual 2:3 MMR and the disk's viscosity and aspect ratio
are sufficiently small, the planets do not migrate toward the Sun.
The mechanism is robust with respect to grid resolution used for the
disk, the smoothing length used for the regularization of the
gravitational potential, and the accretion of mass onto the
planets. In most cases, the planets migrate outward, which is not a
viable evolution in our solar system, because it would imply that
Jupiter was in the asteroid belt in the past.  However, there is a
range of values of viscosity and disk's scale height such that, once
in resonance, the planets have a quasi-stationary evolution during
which their semi major axes remain practically constant. We argue that
Jupiter and Saturn actually followed this kind of evolution.

In general terms for a pair of planets, a quasi-stationary solution can be
found only if the outer planet is significantly less massive than the inner
one, and if the planets are locked in a resonance characterized by a small
orbital separation, so that the gaps opened by the planets in the disk can
overlap. We find that these conditions are not satisfied by any known
extra-solar system of multiple planets. This is consistent with these planets
having suffered a significant migration, that brought them close to the parent
star where they could be discovered. We predict that systems similar to the
Jupiter-Saturn case in terms of mass ratio and separation will be discovered
only when it will be possible to detect distant planets that did not migrate
substantially.

The results of this paper support the view, proposed in Tsiganis et al. (2005)
and Gomes et al. (2005), that the giant planets of the solar system, at the
end of the gas disk era, were on orbits with small mutual separation. However,
from the quantitative point of view, supporting the initial conditions adopted
in the model of Tsiganis et al. and Gomes et al. is problematic. We suggest
that a late fast growth of Saturn's mass or, more likely, a late
enhancement of the viscosity towards the end of the disk's lifetime, could
have extracted Saturn from the 2:3 resonance with Jupiter and driven it close
to the 1:2 resonance. We supported this scenario with simulations, but which
are nevertheless qualitative, given our limited knowledge of process of planet
growth and of disk disappearance.

\acknowledgments

We are grateful to Frederic Masset for his suggestions
and for a careful reading of this manuscript. We also thank the two
anonymous reviewers for their constructive suggestions. We are grateful to the
National Program of Planetology for support.

\centerline\textbf{ REFERENCES}

\begin{itemize}

\item[] Balbus, S.~A., 
Hawley, J.~F.\ 1991.\ A powerful local shear instability in weakly 
magnetized disks. I - Linear analysis. II - Nonlinear evolution.\ 
Astrophysical Journal 376, 214-233. 

\item[] Bate, M.~R., Lubow, S.~H., 
Ogilvie, G.~I., Miller, K.~A.\ 2003.\ Three-dimensional calculations of 
high- and low-mass planets embedded in protoplanetary discs.\ Monthly 
Notices of the Royal Astronomical Society 341, 213-229. 

\item[] Crida, A., Morbidelli, 
A., Masset, F.\ 2006.\ On the width and shape of gaps in protoplanetary 
disks.\ Icarus 181, 587-604. 

\item[] Crida, A., Morbidelli, A., Masset, F. 2007. Simulating planet migration in globally evolving 
disks.\ Astronomy and Astrophysics 461, 1173-1183.

\item[] Crida, A., Morbidelli, A., 2007. Cavity opening by a giant
planet in a protoplanetary disk and effects on planetary
migration. MNRAS, under revision.

%\item[] Duncan, M.~J., Levison, 
%H.~F., Lee, M.~H.\ 1998.\ A Multiple Time Step Symplectic Algorithm for 
%Integrating Close Encounters.\ Astronomical Journal 116, 2067-2077. 

\item[] Gammie, C.~F.\ 1996.\ Layered 
Accretion in T Tauri Disks.\ Astrophysical Journal 457, 355. 

\item[] Goldreich, P., 
Tremaine, S.\ 1980.\ Disk-satellite interactions.\ Astrophysical Journal 
241, 425-441. 

\item[] Gomes, R., Levison, 
H.~F., Tsiganis, K., Morbidelli, A.\ 2005.\ Origin of the cataclysmic Late 
Heavy Bombardment period of the terrestrial planets.\ Nature 435, 466-469. 

\item[] Guillot, T., Hueso, 
R.\ 2006.\ The composition of Jupiter: sign of a (relatively) late 
formation in a chemically evolved protosolar disc.\ Monthly Notices of the 
Royal Astronomical Society 367, L47-L51. 

\item[] Haghighipour, 
N., Boss, A.~P.\ 2003.\ On Gas Drag-Induced Rapid Migration of Solids in a 
Nonuniform Solar Nebula.\ Astrophysical Journal 598, 1301-1311. 

\item[] Hayashi C. 1981. Structure of the solar nebula, growth and decay of magnetic
fields and effects of magnetic and turbulent viscosities on the
nebula. {\it Prog. Theor. Phys. Suppl.}, {\bf 70}, 35-53.

\item[] Ilgner, M., Nelson, 
R.~P.\ 2006a.\ On the ionisation fraction in protoplanetary disks. I. 
Comparing different reaction networks.\ Astronomy and Astrophysics 445, 
205-222. 

Ilgner, M., Nelson, 
R.~P.\ 2006b.\ On the ionisation fraction in protoplanetary disks. II. The 
effect of turbulent mixing on gas-phase chemistry.\ Astronomy and 
Astrophysics 445, 223-232. 

\item[] Kley, W.\ 1999.\ Mass flow and 
accretion through gaps in accretion discs.\ Monthly Notices of the Royal 
Astronomical Society 303, 696-710. 

\item[] Kley, W., D'Angelo, G., 
Henning, T.\ 2001.\ Three-dimensional Simulations of a Planet Embedded in a 
Protoplanetary Disk.\ Astrophysical Journal 547, 457-464. 

\item[] Levison H.~F., Morbidelli, A., Gomes, R. and Tsiganis, K. 2007
Origin of the structure of the Kuiper Belt during {a
    Dynamical Instability in the Orbits of Uranus and
    Neptune}. Icarus, submitted.

\item[] Lin, D.~N.~C., 
Papaloizou, J.\ 1979.\ Tidal torques on accretion discs in binary systems 
with extreme mass ratios.\ Monthly Notices of the Royal Astronomical 
Society 186, 799-812. 

\item[] Lin, D.~N.~C., 
Papaloizou, J.\ 1986.\ On the tidal interaction between protoplanets and 
the protoplanetary disk. III - Orbital migration of protoplanets.\ 
Astrophysical Journal 309, 846-857. 

\item[] Lubow, S.~H., Seibert, 
M., Artymowicz, P.\ 1999.\ Disk Accretion onto High-Mass Planets.\ 
Astrophysical Journal 526, 1001-1012. 

\item[] Lynden-Bell, 
D., Pringle, J.~E.\ 1974.\ The evolution of viscous discs and the origin of 
the nebular variables..\ Monthly Notices of the Royal Astronomical Society 
168, 603-637. 

\item[] Malhotra, R.\ 1993.\ The 
Origin of Pluto's Peculiar Orbit.\ Nature 365, 819. 

\item[] Masset, F.\ 2000a.\ FARGO: A 
fast eulerian transport algorithm for differentially rotating disks.\ 
Astronomy and Astrophysics Supplement Series 141, 165-173.

\item[] Masset, F.~S.\ 2000b.\ FARGO: A 
Fast Eulerian Transport Algorithm for Differentially Rotating Disks.\ ASP 
Conf.~Ser.~219: Disks, Planetesimals, and Planets 219, 75. 

%\item[] Masset, F.~S.\ 2001.\ On the 
%Co-orbital Corotation Torque in a Viscous Disk and Its Impact on Planetary 
%Migration.\ Astrophysical Journal 558, 453-462. 

\item[] Masset, F., 
Snellgrove, M.\ 2001.\ Reversing type II migration: resonance trapping of a 
lighter giant protoplanet.\ Monthly Notices of the Royal Astronomical 
Society 320, L55-L59. 

\item[] Masset, F.~S., 
Papaloizou, J.~C.~B.\ 2003.\ Runaway Migration and the Formation of Hot 
Jupiters.\ Astrophysical Journal 588, 494-508. 

\item[] Masset, F.~S., 
Morbidelli, A., Crida, A., Ferreira, J.\ 2006.\ Disk Surface Density 
Transitions as Protoplanet Traps.\ Astrophysical Journal 642, 478-487. 

\item[] Masset, F.~S. 2006. Planet-disk interactions. In Tidal
Interactions in Composite Systems M.-J. Goupil and J.-P. Zahn (eds)
EAS Publications Series,in press.

\item[] Morbidelli, A., 
Levison, H.~F., Tsiganis, K., Gomes, R.\ 2005.\ Chaotic capture of 
Jupiter's Trojan asteroids in the early Solar System.\ Nature 435, 462-465.

\item[] Papaloizou, J., 
Lin, D.~N.~C.\ 1984.\ On the tidal interaction between protoplanets and the 
primordial solar nebula. I - Linear calculation of the role of angular 
momentum exchange.\ Astrophysical Journal 285, 818-834. 

\item[] Pollack, J.~B., 
Hubickyj, O., Bodenheimer, P., Lissauer, J.~J., Podolak, M., Greenzweig, 
Y.\ 1996.\ Formation of the Giant Planets by Concurrent Accretion of Solids 
and Gas.\ Icarus 124, 62-85. 

\item[] Rice, W.~K.~M., Wood, K., 
Armitage, P.~J., Whitney, B.~A., Bjorkman, J.~E.\ 2003.\ Constraints on a 
planetary origin for the gap in the protoplanetary disc of GM Aurigae.\ 
Monthly Notices of the Royal Astronomical Society 342, 79-85. 

\item[] Shakura, N.~I., 
Sunyaev, R.~A.\ 1973.\ Black holes in binary systems. Observational 
appearance..\ Astronomy and Astrophysics 24, 337-355. 

\item[] Thommes, E.~W.\ 2005.\ A 
Safety Net for Fast Migrators: Interactions between Gap-opening and 
Sub-Gap-opening Bodies in a Protoplanetary Disk.\ Astrophysical Journal 
626, 1033-1044. 

\item[] Tsiganis, K., Gomes, 
R., Morbidelli, A., Levison, H.~F.\ 2005.\ Origin of the orbital 
architecture of the giant planets of the Solar System.\ Nature 435, 
459-461. 

\item[] Varni{\`e}re, P., 
Blackman, E.~G., Frank, A., Quillen, A.~C.\ 2006.\ Planets Rapidly Create 
Holes in Young Circumstellar Disks.\ Astrophysical Journal 640, 1110-1114. 

\item[] Ward, W.~R.\ 1986.\ Density waves in the solar nebula -
  Differential Lindblad torque.\ Icarus 67, 164-180.

\item[] Ward, W.~R.\ 1997.\ Protoplanet 
Migration by Nebula Tides.\ Icarus 126, 261-281. 
\end{itemize}

\newpage

\centerline{\bf Figure captions}

\begin{figure}[t!]
\centerline{\psfig{figure=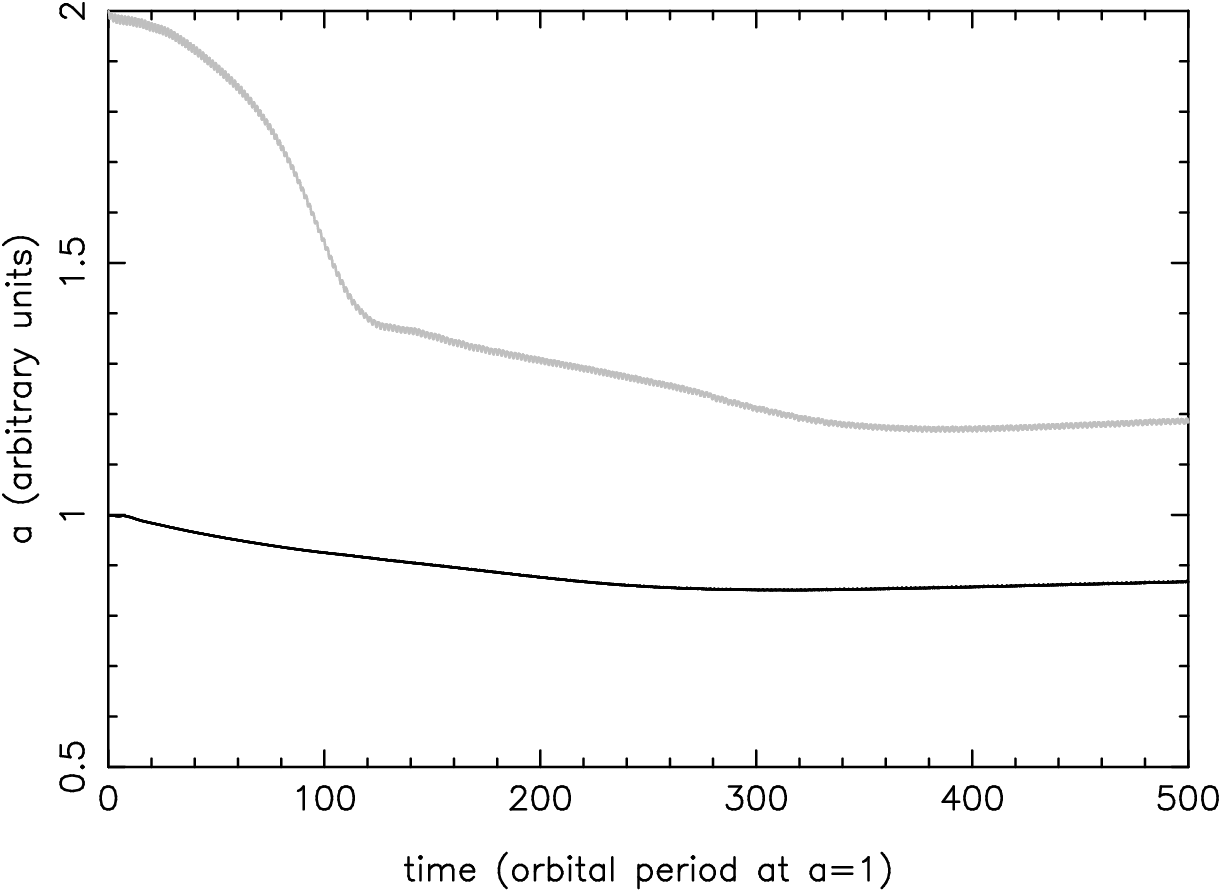,height=14.cm}}
\vspace*{-.3cm}
\caption{}
\label{MS-orig}
\end{figure}

\begin{itemize}

\item[Fig. 1] An illustration of the dynamical evolution described in MS01. The
  black and grey curves show the evolutions of the semi major axes of Jupiter
  and Saturn respectively. Capture in the 2:3 MMR occurs when the migration of
  Saturn is reversed.

\item[Fig. 2] The same as Fig.~\ref{MS-orig}, but for an initial
  location of Saturn close (but not into) the 2:3 MMR with Jupiter.

\item[Fig. 3] The evolution of Jupiter (lower set of curves starting at $r=1$)
  and Saturn (upper set of curves, starting at $r=1.4$).  Different grey
  levels refer to different aspect ratios, as labeled. The viscosity is
  independent of radius and equal to $10^{-5.5}$ in all cases. The planets are
  released after 400 orbits.

\item[Fig. 4] The dash-dotted curve shows the initial density profile
of the disk, adopted in all simulations. The solid curves show the
density profile corresponding to the moment when the planets are
released, at 1 and 1.4 respectively for Jupiter and Saturn. Different
grey levels refer to different aspect ratios, as labeled. The
viscosity is independent of radius and equal to $10^{-5.5}$ in all
cases.

\item[Fig. 5] {\bf The evolution of Jupiter and Saturn in two simulations,
  both with $H/r$=3\% and $\nu=10^{-5.5}$. The black curves refer to
  the simulation adopting the initial disk surface density shown in
  Fig.~\ref{smalldisk}. The grey curves adopt an initial surface density
  profile that has been divided by a factor of 2. The planets are
  released in the two cases after about 400 orbital periods at $r=1$. 
  The gray curves are plotted after rescaling the time as
  $t'=(t-400)/2$. The fact that gray and balck curves overlap shows
  that the evolution is the same, but the migration speed is reduced
  proportionally with the mass of the disk.}   

\item[Fig. 6] The surface density of the disk in the simulation with
$H/r=3$\%.  Different grey levels refer to different times, labeled in
unit of initial Jupiter's orbital period. The planets are released at
time $T=400$.

\item[Fig. 7] The evolution of Jupiter and Saturn in three simulations, with
  $H/r=3$\% and $\nu=10^{-5.5}$. The simulation plotted in black and labeled
  `Hill' is the one already shown in Fig.~\ref{MS-Hr}. The simulation
  reported in grey and labeled `H' adopts a different prescription for the
  smoothing length, which is now imposed equal to 70\% of the local thickness
  of the disk. The simulation plotted in light grey and labeled 'High res.'
  is the same as the latter simulation, but with azimuthal and radial grid
  resolutions increased by a factor of 2. Time is measured relative to the
  instant $T_0$ when Saturn starts to migrate outward. The semi major axes of
  the planets are normalized relative to the semi major axis of Jupiter at
  $T_0$. This allows a more direct comparison among the three
  simulations.

\item[Fig. 8] {\bf The evolution of Jupiter and Saturn in two simulations,
  with $H/r=4$\%, $\nu=10^{-5.5}$. The simulation represented by black
  curves, labelled `0.6 Hill', is the one already shown in
  Fig.~\ref{MS-Hr}, obtained adopting a smoothing length $\rho=0.6
  R_H$. That represented by grey curves, labelled `0.25 Hill' has been
  obtained with $\rho=0.25 R_H$. Finally, the simulation represented
  by light grey curves, labelled `Excl. Hill' has also been obtained
  with $\rho=0.25 R_H$, but excluding the torques exerted on the
  planets by the gas inside their Hill spheres.} 

\item[Fig. 9] {\bf The evolution of Jupiter and Saturn in two simulations, with
  $H/r=4$\%, $\nu=10^{-5.5}$ and $\rho=0.7 R_H$. 
  In the simulation plotted in black and labeled
  `Nominal', the planets have been released on free-to-evolve orbits after 400
  orbital periods at $r=1$. In the simulation
  reported in grey and labeled `Late release' the planets have been
  released after 5,000 orbits. Notice that the grey curves have been
  shifted downwards by 1\%, in order to avoid a perfect overlap with
  the black curves. Therefore,  evolution of the planets after the release
  time is the same.}

\item[Fig. 10] The evolution of Jupiter (lower set of curves starting at $r=1$)
  and Saturn (upper set of curves, starting at $r=1.4$).  Different grey
  levels refer to different viscosities, as labeled. The aspect ratio $H/r$ is
  equal to $4$\% in all cases. The planets are released
  after 400 initial Jovian orbits.

\item[Fig. 11] The density profiles of the disk
at the moment when the planets are released. Different grey levels refer to
different viscosities, as labeled. The aspect ratio is 4\% in all
simulations.

\item[Fig. 12] The grey curves show the evolutions of Saturn (starting
  at 1.4) and Jupiter (starting at 1) in the simulation with
  $H/r=3$\%, $\nu=10^{-5.5}$ and $\rho=0.7 H$. Time is counted in
  initial Jovian orbital periods, from the instant when the planets
  are released. The black curves show how the evolutions of Saturn and
  Jupiter change, starting from $t=900$, in the case where Saturn (but
  not Jupiter) is allowed to accrete mass with a removal rate of 1.

\item[Fig. 13] The evolution of two planets with 3 Jupiter masses
  (starting at $r=1$) and 1 Jupiter masses (starting at $r=1.4$). The
  disk aspect ratio is 3\% and the viscosity is $10^{-5.5}$,
  independent of radius. A smoothing length equal to 70\% of the local
  disk's height is used.

\item[Fig. 14] The black solid curves show the evolution of Jupiter and Saturn.
  As usual, the planets start at $r=1$ and 1.4 respectively, and are released
  after 400 initial Jovian orbital periods. The disk scale heigh is 5\% and
  the viscosity is $10^{-5.5}$. After that the planets lock in the 2:3 MMR (at
  $t\sim 900$ the planets' semi major axes remain substantially constant.  The
  grey curves show the location  of the 2:3 and 1:2 MMR with
  Jupiter, according to Kepler law. Notice that the semi major axis of Saturn
  is slightly larger than that corresponding to the 2:3 MMR in the Kepler
  approximation, due to the effects of the disk's gravity. At $t=3200$ (marked
  by a vertical dashed line), the viscosity of the disk is increased at a rate
  of $6.28\times 10^{-9}$ per orbital period. This eventually forces Jupiter
  to migrate inward and extracts Saturn from the 2:3 MMR. The simulation is
  stopped when Saturn reaches the vicinity of the 1:2 MMR with
  Jupiter.

\end{itemize}

\newpage

\newpage
\begin{figure}[t!]
\centerline{\psfig{figure=I09908_Fig2.ps,height=14.cm}}
\vspace*{-.3cm}
\caption{}
\label{MS-proches}
\end{figure} 

\newpage
\begin{figure}[t!]
\centerline{\psfig{figure=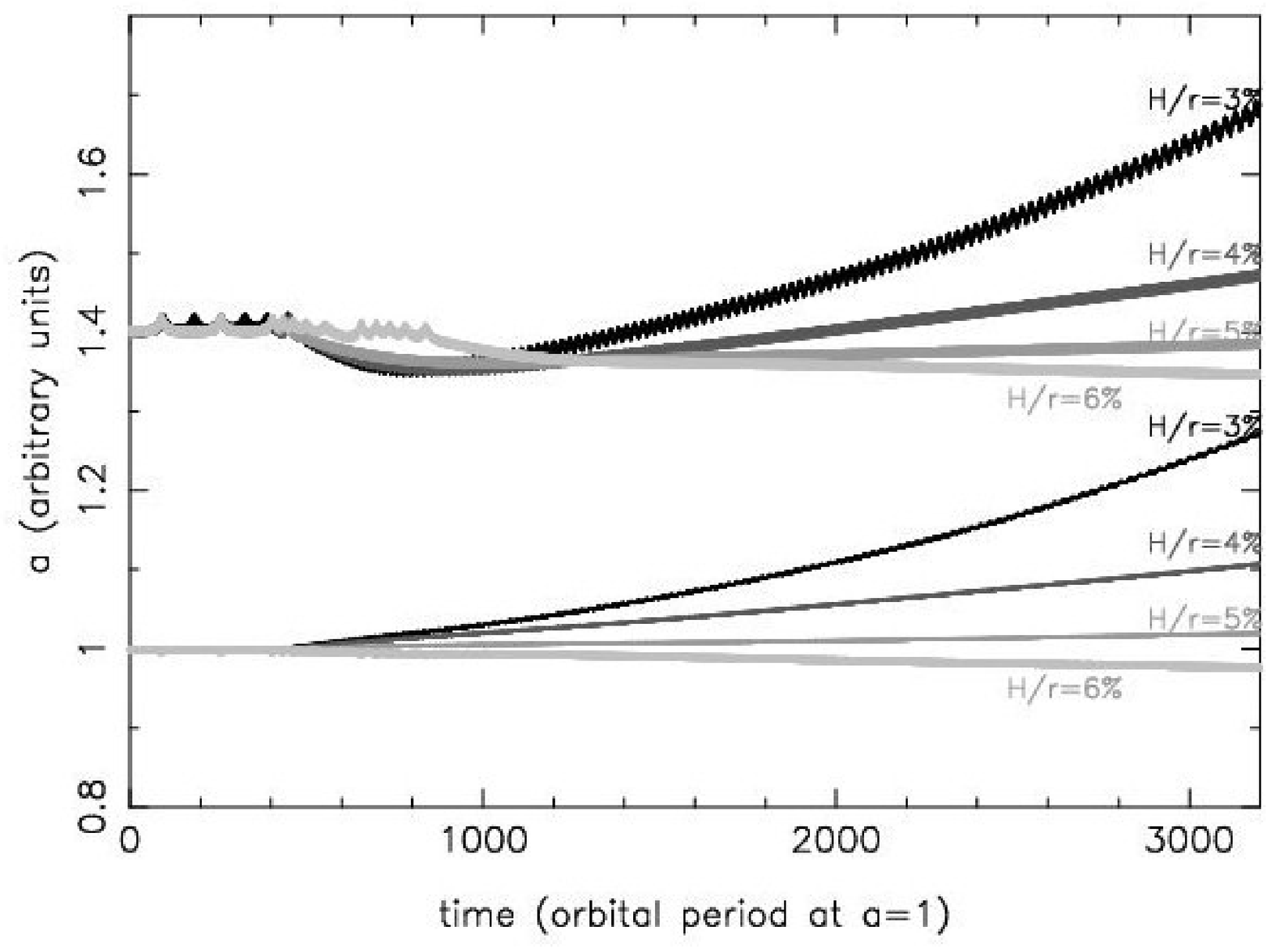,height=14.cm}}
\vspace*{-.3cm}
\caption{}
\label{MS-Hr}
\end{figure} 

\newpage
\begin{figure}[t!]
\centerline{\psfig{figure=I09908_Fig4.ps,height=14.cm}}
\vspace*{-.3cm}
\caption{}
\label{profiles}
\end{figure} 

\newpage
\begin{figure}[t!]
\centerline{\psfig{figure=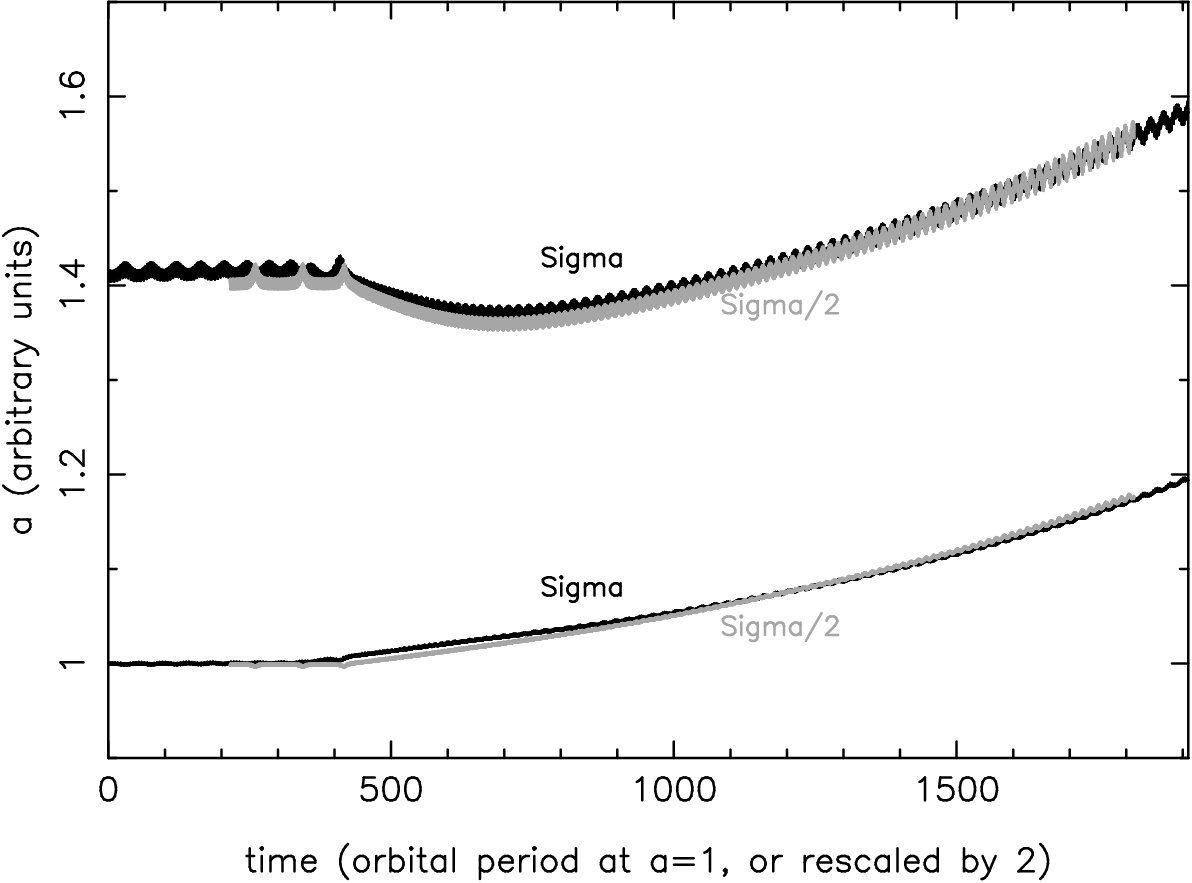,height=14.cm}}
\vspace*{-.3cm}
\caption{}
\label{smalldisk}
\end{figure} 

\newpage
\begin{figure}[t!]
\centerline{\psfig{figure=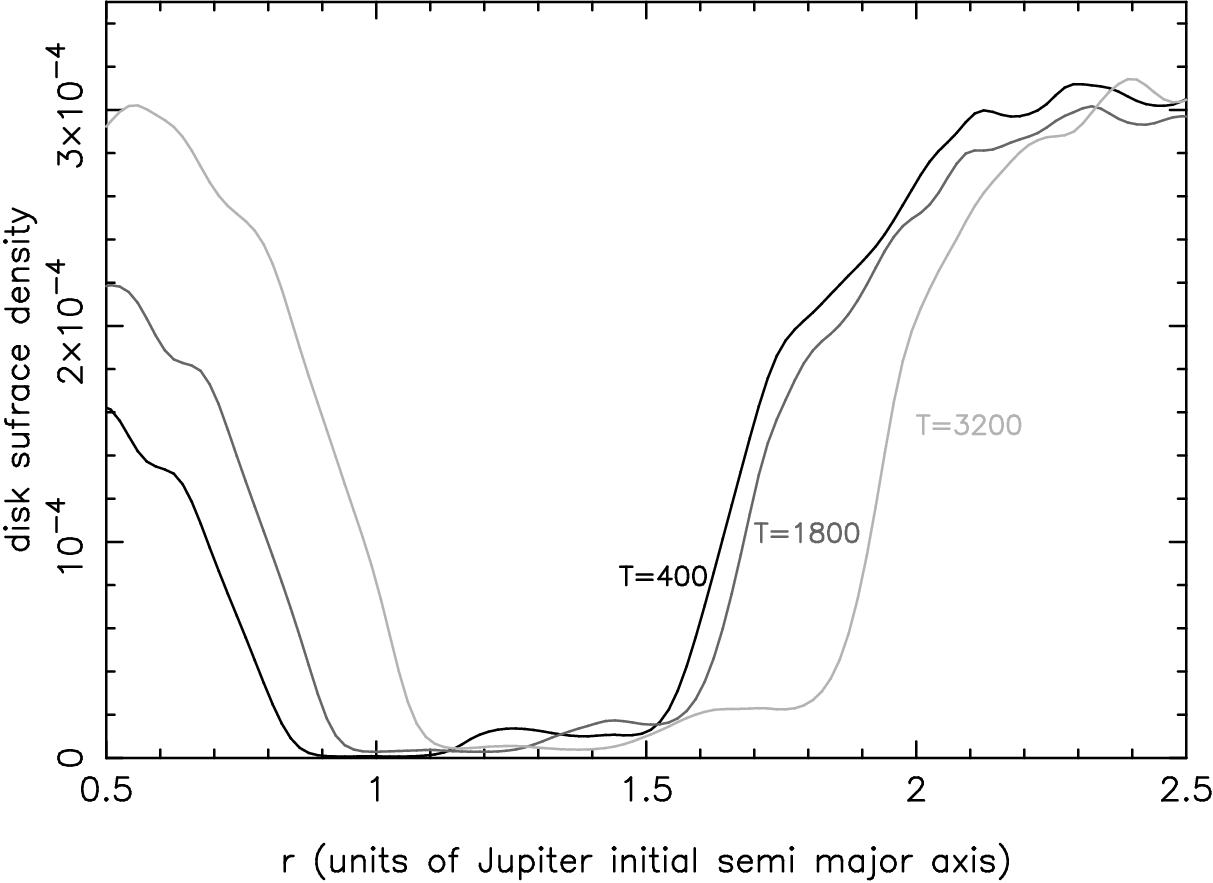,height=14.cm}}
\vspace*{-.3cm}
\caption{}
\label{profiles_t}
\end{figure} 

\newpage
\begin{figure}[t!]
\centerline{\psfig{figure=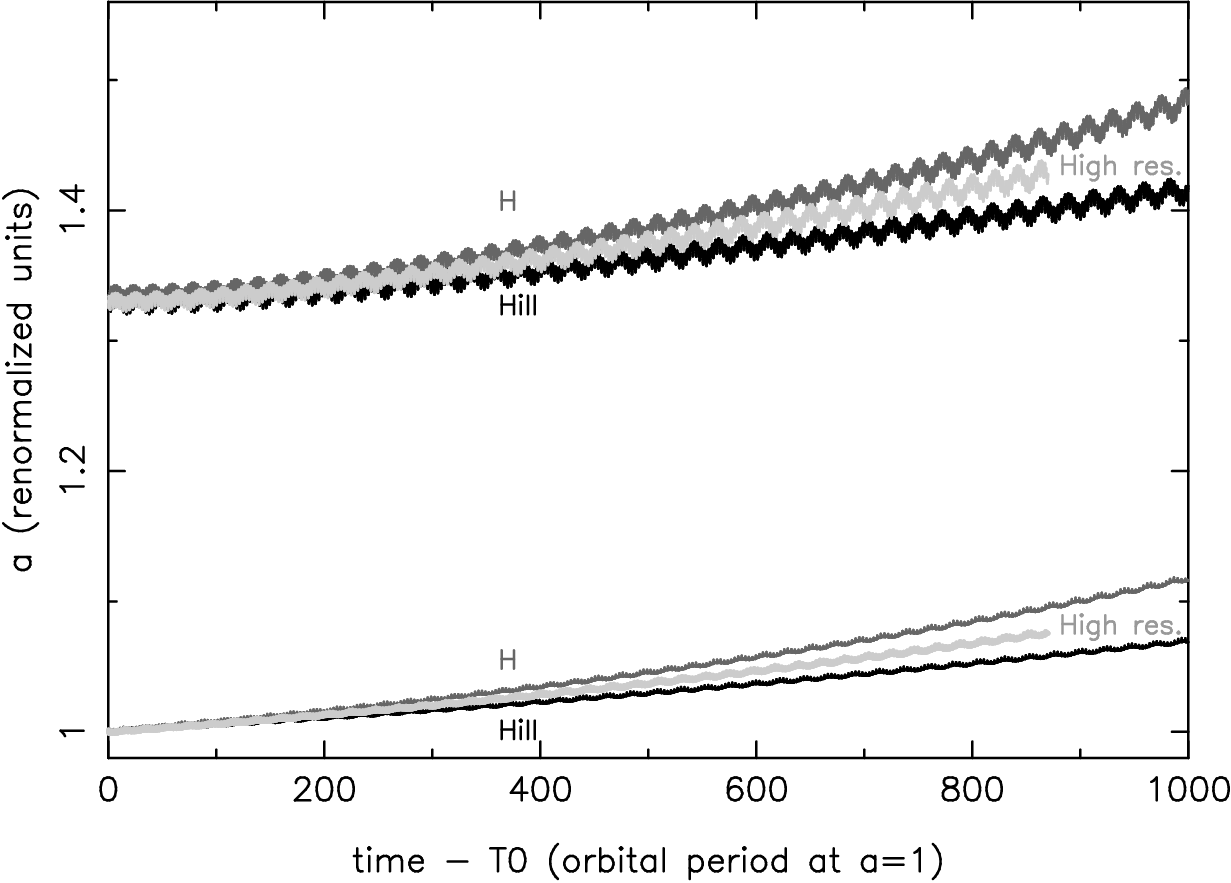,height=14.cm}}
\vspace*{-.3cm}
\caption{}
\label{comp-tch}
\end{figure} 

\newpage
\begin{figure}[t!]
\centerline{\psfig{figure=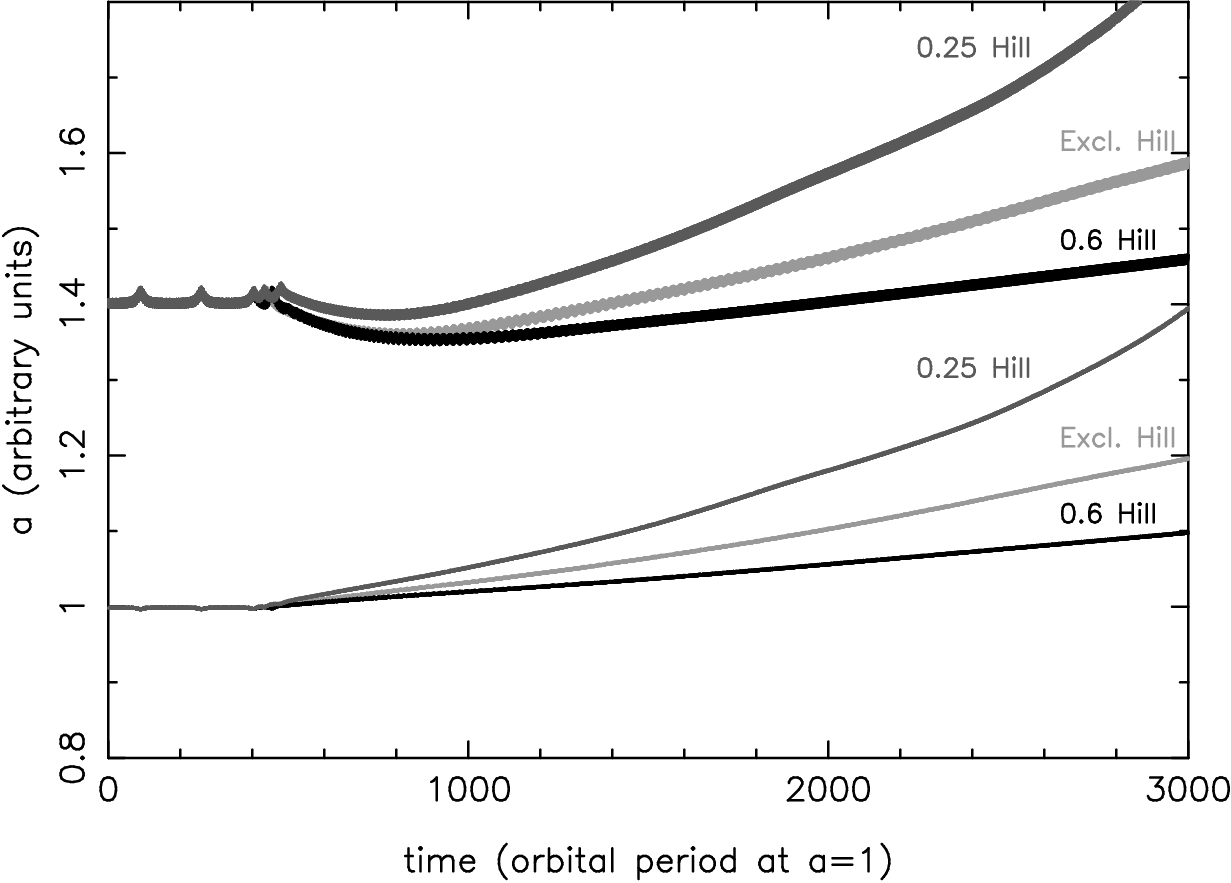,height=14.cm}}
\vspace*{-.3cm}
\caption{}
\label{short-smooth}
\end{figure} 

\newpage
\begin{figure}[t!]
\centerline{\psfig{figure=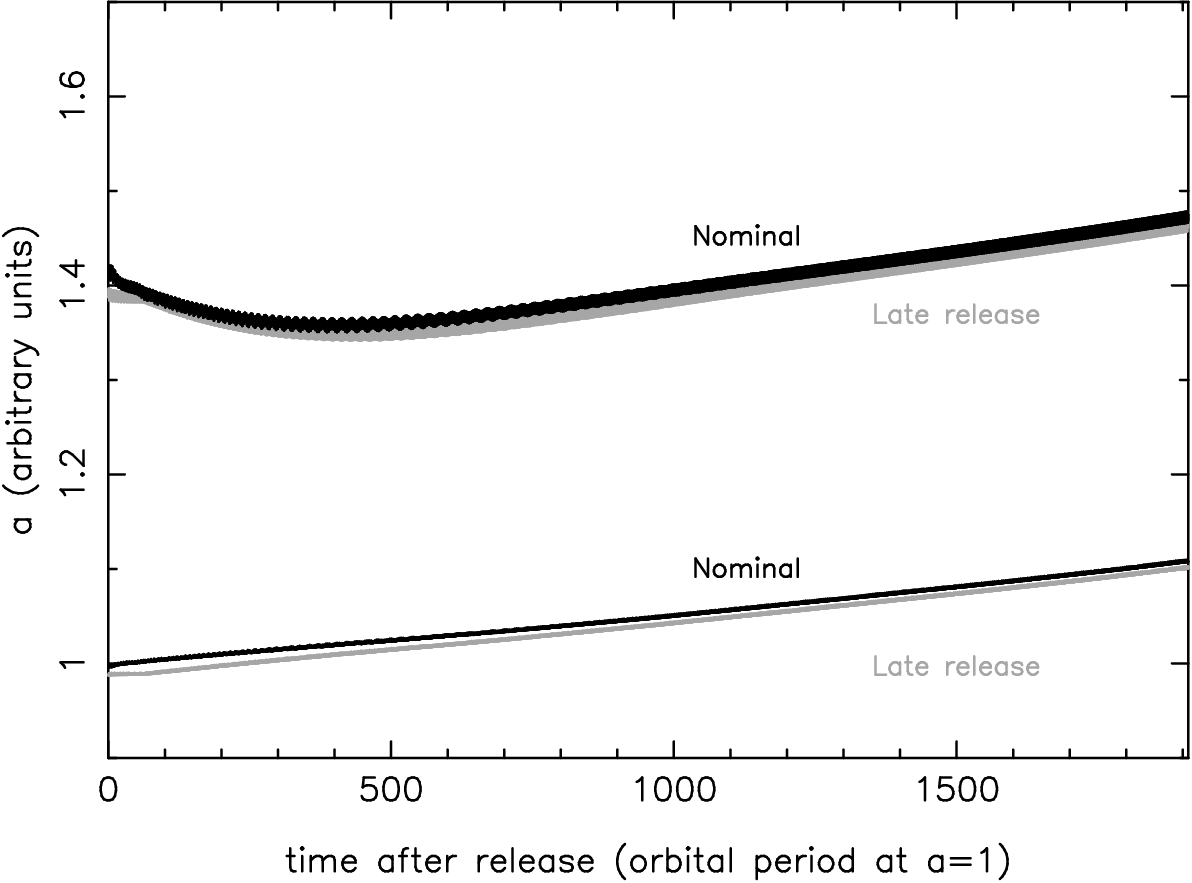,height=14.cm}}
\vspace*{-.3cm}
\caption{}
\label{release}
\end{figure} 

\newpage
\begin{figure}[t!]
\centerline{\psfig{figure=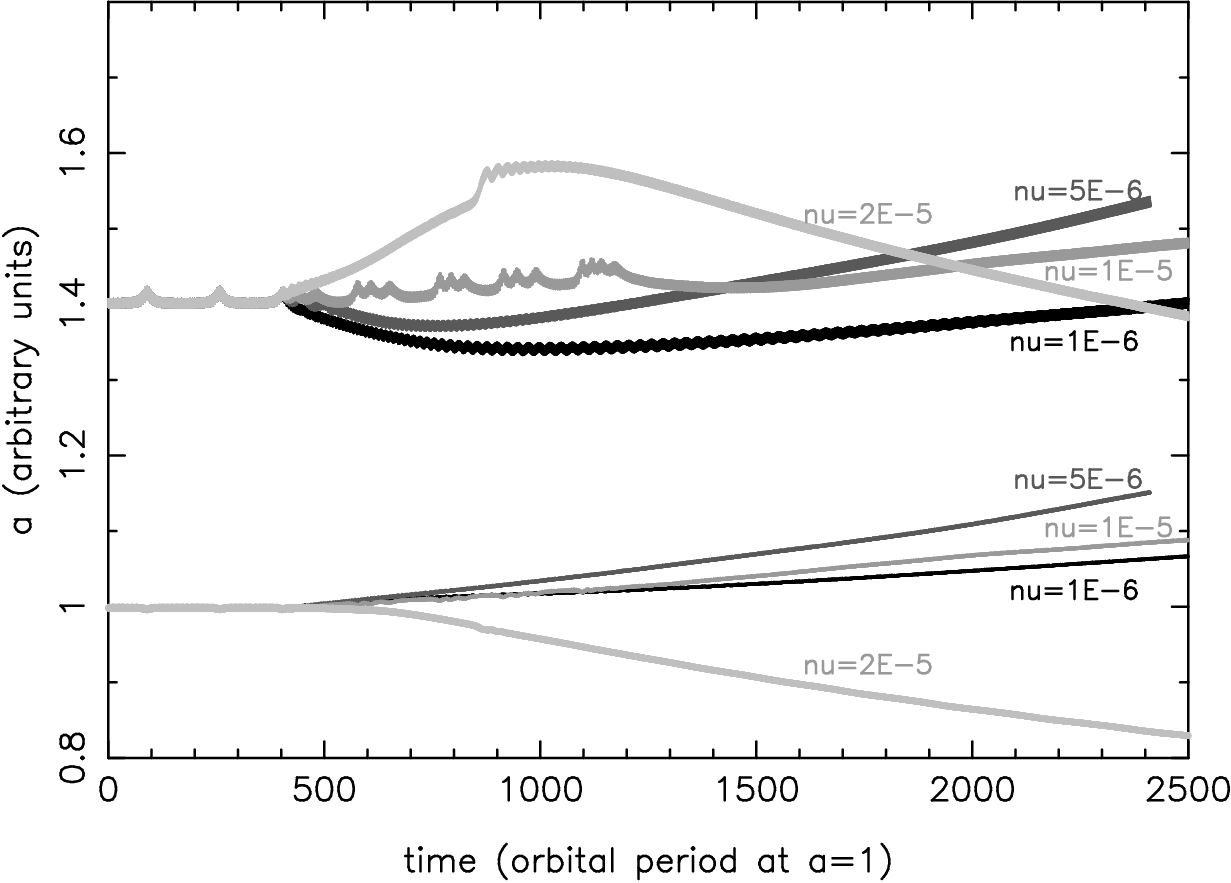,height=14.cm}}
\vspace*{-.3cm}
\caption{}
\label{MS-nu}
\end{figure} 

\newpage
\begin{figure}[t!]
\centerline{\psfig{figure=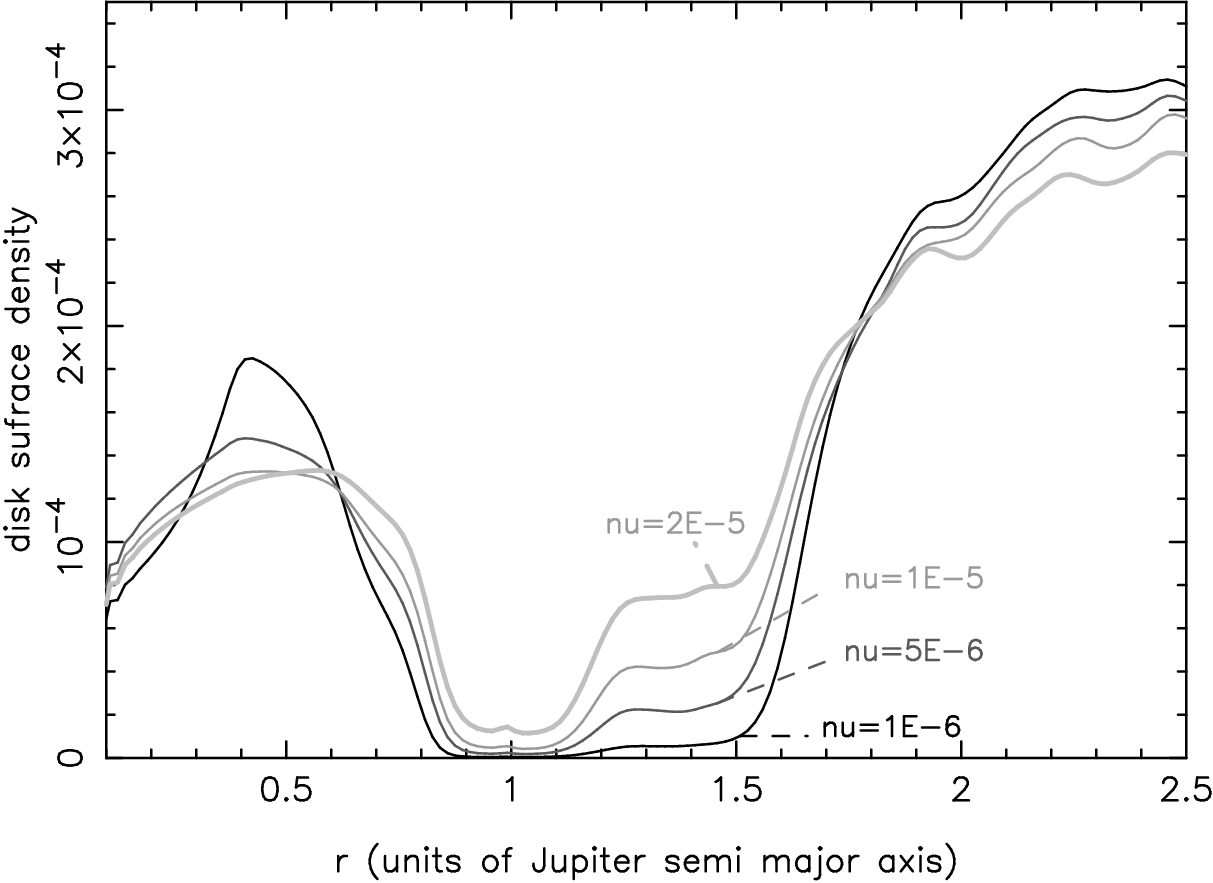,height=14.cm}}
\vspace*{-.3cm}
\caption{}
\label{profiles2}
\end{figure} 

\newpage
\begin{figure}[t!]
\centerline{\psfig{figure=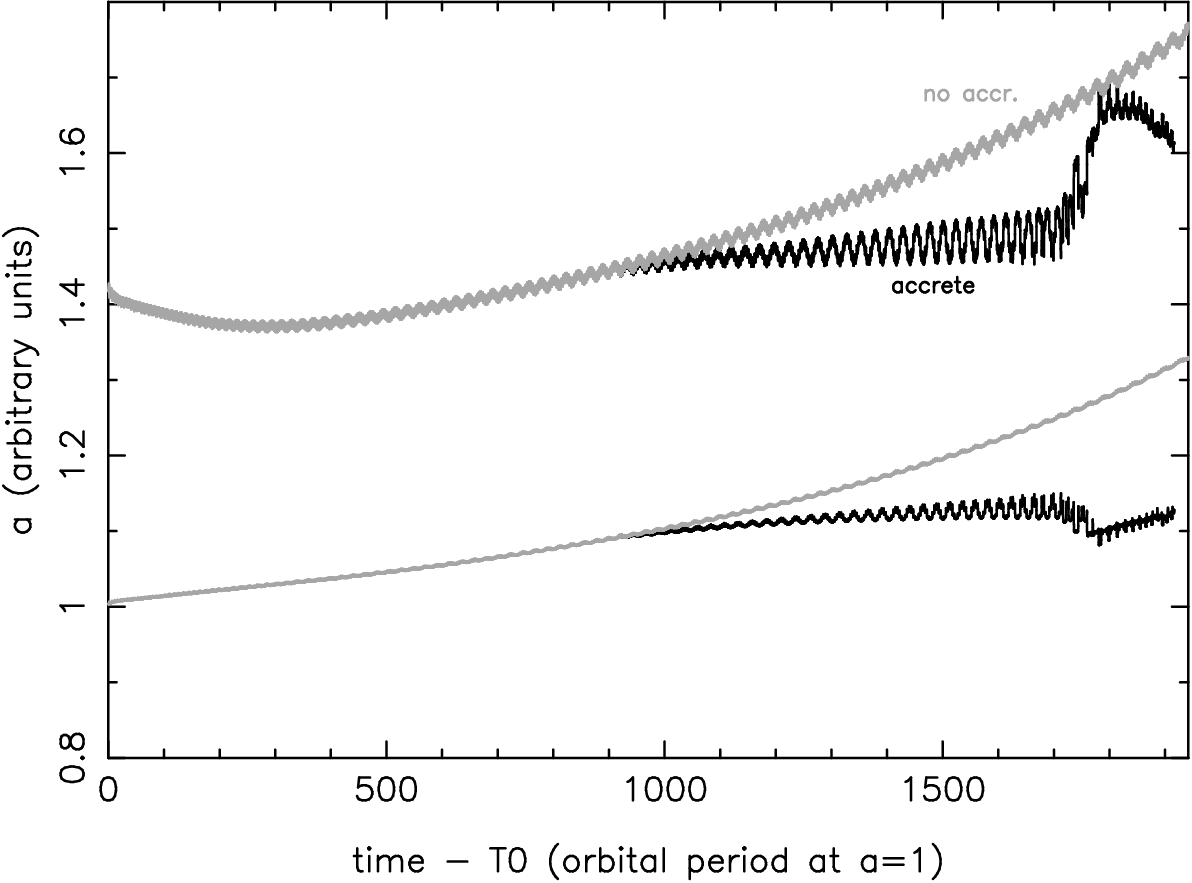,height=14.cm}}
\vspace*{-.3cm}
\caption{}
\label{MS-accr}
\end{figure} 

\newpage
\begin{figure}[t!]
\centerline{\psfig{figure=I09908_Fig13.ps,height=14.cm}}
\vspace*{-.3cm}
\caption{}
\label{3J-J}
\end{figure} 

\newpage
\begin{figure}[t!]
\centerline{\psfig{figure=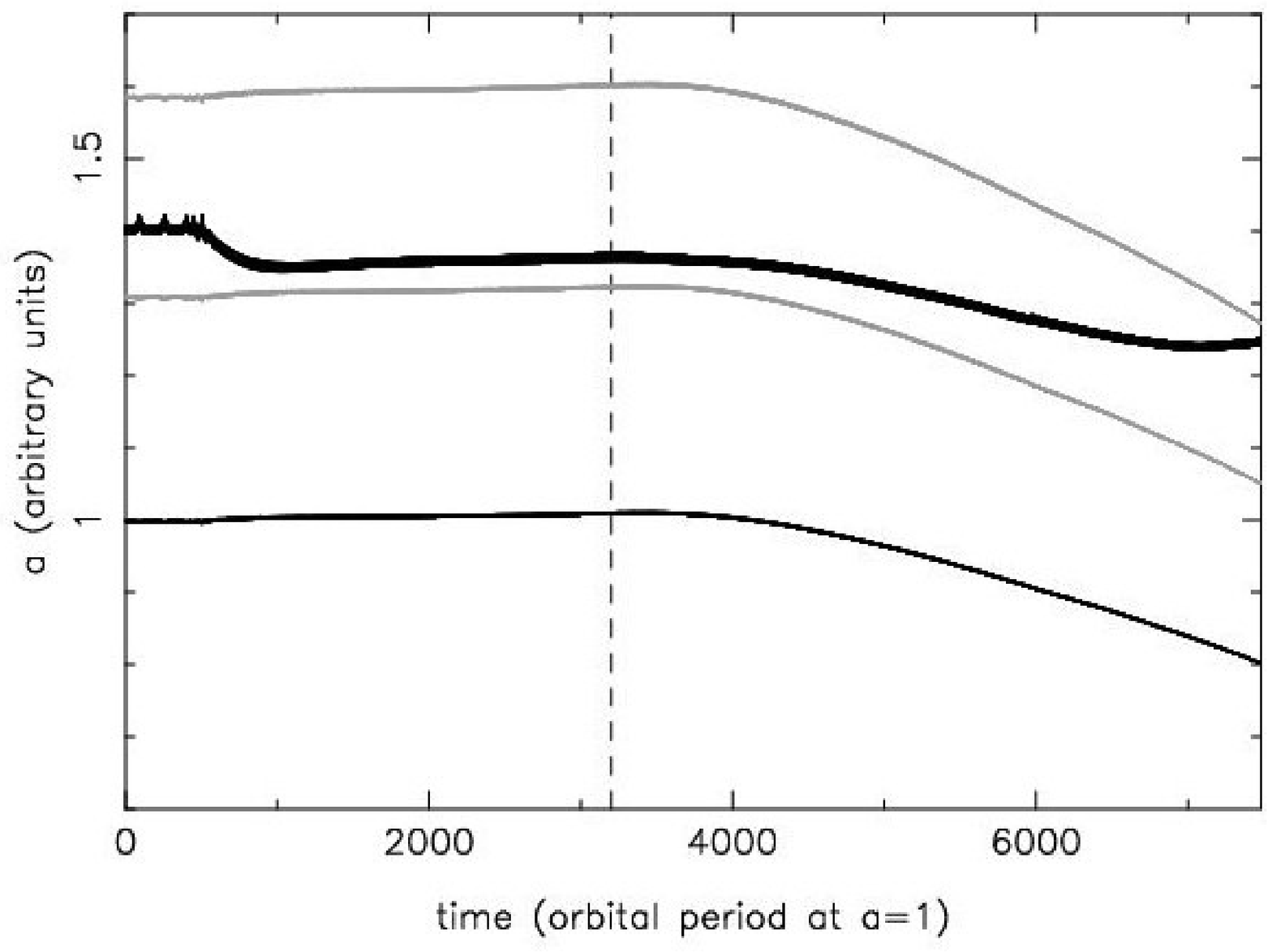,height=14.cm}}
  \vspace*{-.3cm}
\caption{}
\label{varnu}
\end{figure}

\end{document}